\documentclass[preprint,12pt]{elsarticle}

\usepackage{amssymb}
\usepackage{subcaption}

\usepackage{tcolorbox}
\usepackage{multicol}
\usepackage{booktabs}
\usepackage{hyperref}
\usepackage{array}

\usepackage{graphicx} 
\usepackage{colortbl} 
\journal{}

\begin{document}

\begin{frontmatter}






\title{Hackphyr: A Local Fine-Tuned LLM Agent for Network Security Environments}


\author[inst1]{Maria Rigaki}

\affiliation[inst1]{organization={Faculty of Electrical Engineering, Czech Technical University in Prague},
            city={Prague},
            country={Czech Republic}}

\author[inst2,inst3]{Carlos Catania}
\author[inst1]{Sebastian Garcia}

\affiliation[inst2]{organization={Engineering School, National University of Cuyo},
            city={Mendoza},
            country={Argentina}}

\affiliation[inst3]{organization={National Scientific and Technical Research Council (CONICET)},
            country={Argentina}}

\begin{abstract}

Large Language Models (LLMs) have shown remarkable potential across various domains, including cybersecurity. Using commercial cloud-based LLMs may be undesirable due to privacy concerns, costs, and network connectivity constraints. In this paper, we present Hackphyr, a locally fine-tuned LLM to be used as a red-team agent within network security environments. Our fine-tuned 7 billion parameter model can run on a single GPU card and achieves performance comparable with much larger and more powerful commercial models such as GPT-4. Hackphyr clearly outperforms other models, including GPT-3.5-turbo, and baselines, such as Q-learning agents in complex, previously unseen scenarios. To achieve this performance, we generated a new task-specific cybersecurity dataset to enhance the base model's capabilities. Finally, we conducted a comprehensive analysis of the agents' behaviors that provides insights into the planning abilities and potential shortcomings of such agents, contributing to the broader understanding of LLM-based agents in cybersecurity contexts.

\end{abstract}




\end{frontmatter}


\section{Introduction}
\label{sec:intro}


Autonomous agents are systems that interact with and adapt to their environments to achieve specific goals. Traditionally, human defenders have been the primary responders to security breaches. The potential deployment of Autonomous Intelligent Cyber Agents (AICA)~\cite{kott_autonomous_2023} for defensive operations could transform cyber defenses. At the same time, offensive agents, also known as Red-team Agents, have the potential to test existing defenses and uncover potential problems. These agents could preemptively identify threats making defense a dynamic, continuous process. 


In recent years, Reinforcement Learning (RL) techniques have emerged as the predominant method for developing autonomous agents to navigate and adapt to dynamic environments~\cite{mnih_human-level_2015,ghanem_hierarchical_2023}. Reinforcement learning-based agents evolve through direct interactions with their environments, learning to optimize behaviors through a trial-and-error approach supported by reward systems. Trial-and-error comes with certain shortcomings: it requires hundreds of thousands of training episodes to learn a policy, and once trained, the policy is not flexible. An agent trained in one specific environment configuration will not work without retraining in a slightly different scenario. 


LLMs have achieved notable successes in recent years, demonstrating significant potential in mimicking human-like behavior~\cite{park_generative_2023,xi2023rise}. Building upon this capability, a growing research area has employed LLMs as central controllers to construct autonomous agents to obtain human-like decision-making capabilities~\cite{du_guiding_2023,wang2024voyager}. Compared to RL approaches, LLM-based agents encapsulate a vast spectrum of human knowledge and a richer understanding of the world that enables them to act without needing additional or specific training~\cite{wang2024survey}.
Secondly, pre-trained language models exhibit remarkable adaptability, as they can be easily adapted to new tasks or domains with relatively minimal additional training, offering a significant advantage over other machine learning models that may require extensive retraining~\cite{magister_teaching_2023}.


Prior work has explored the intersection of LLMs, cybersecurity, and sequential decision-making~\cite{rigaki_out_2024}. The authors proposed a novel approach that uses pre-trained LLMs as agents within cybersecurity environments. 
However, the models used were proprietary models running on the cloud. The use of cloud-based language models suffers from several problems. These models are costly to access, limiting accessibility and reproducibility. Furthermore, the models are often unilaterally changed without explanations or updates, raising concerns about their stability and reliability. 

The situation can be even more critical in the context of cyber security. Communicating possibly sensitive information about a company network to a cloud-based proprietary model for additional analysis could be a major threat to the security and privacy of the company. In contrast, open-source models are freely available, promoting accessibility, reproducibility, and privacy. Additionally, many open source models come in smaller parameter sizes that allow inference to be performed in lower-cost hardware in the company premises if required. However, smaller model sizes often mean lower performance and the need for further model fine-tuning to a specific task.

In this work, we fine-tuned a 7 billion pre-trained LLM (Zephyr-7b-$\beta$) to be used as a red team agent in the NetSecGame environment~\cite{rigaki_out_2024}. For the fine-tuning process, we created a new dataset that helped boost the base model's understanding of the environment and its ability to propose valid and useful actions. Our agent was evaluated in three scenarios with varying levels of complexity and compared to pre-trained commercial LLMs and RL baselines such as Q-learning. Finally, we performed a detailed behavioral analysis of each agent's actions to evaluate their planning ability and how it matches human-like behavior.



This paper makes several key contributions to the field: 
\begin{itemize}
    \item A locally deployed model named Hackphyr, specifically fine-tuned to work in the NetSecGame environment.

    \item A novel dataset, accompanied by an ablation study that analyzes the impact of different dataset components on model performance. This analysis provides valuable insights into the factors that most significantly influence the effectiveness of the fine-tuned model in network security environments.

    \item A detailed behavioral analysis of three LLM-based agents (GPT-4, Zephyr-7b-$\beta$, Hackphyr). This analysis offers a unique perspective on the decision-making processes of each model, providing information on their ability to navigate and solve network security challenges autonomously.
\end{itemize}

The rest of the paper is organized as follows: Sections~\ref{sec:background} and~\ref{sec:related_work} present the background knowledge and related work, respectively. The NetSecGame environment is presented in Section~\ref{sec:netsec_env}, and the general agent design is described in Section~\ref{sec:agent_design}. Section~\ref{sec:methodology} details the supervised fine-tuning methodology, including the dataset creation and the hyper-parameter tuning process. The design of experiments and the three different scenario configurations are presented in Section~\ref{sec:experiment_design}. The experiments' results and the agents' behavioral analysis are described in Sections~\ref{sec:results} and~\ref{sec:behavioral_analysis}, respectively.
The dataset ablation study details are in Section~\ref{sec:ablation}. Finally, we discuss some findings and limitations of the work in Section~\ref{sec:discussion} and conclude the paper with Section~\ref{sec:conclusions}.

\section{Background}
\label{sec:background}

The term Large Language Models (LLMs) typically refers to transformer-based models~\cite{attention2017} of billions of parameters that are trained to perform tasks related to Natural Language Processing (NLP). The input text is usually split into smaller chunks (tokens), and the models are trained to predict the next token $t$ after the input sequence.

Formally, given a series of tokens $t_1, \dots, t_{N-1}$, a language modelling aims to calculate the joint probability $$P(t_{1} \dots t_{N}) = \prod_{i=1}^{N} P(t_{i} | t_{1}, \dots, t_{N-1})$$ 
Large language models use neural networks (transformers) to estimate this probability and sample the next token in the sequence ${t}_{N} \sim f_{\theta}(t_{N} | t_1, \dots, t_{N-1})$, where $f_{\theta}$ is the trained model with parameters $\theta$. 

There are several stages in the training of LLMs. The first stage is the \textit{unsupervised pre-training} with many tokens. The resulting models can perform next-token prediction and generally perform well in varied tasks related to the data. However, these large models, often called foundational, sometimes do not perform well in certain specialized tasks. Foundational models can adapt to tasks such as classification using Supervised Fine-Tuning (SFT)~\cite{scaling2024,wei2022finetuned}. Fine-tuning requires labeled data, but these datasets are usually much smaller than those used for the pre-training phase. Sometimes, the dataset can be produced using the answers of a more capable model that is used as a teacher. This method is also called distilled SFT (dSFT)~\cite{taori2023alpaca,tunstall_zephyr_2023}. A common fine-tuning use case is \textit{instruction fine-tuning}~\cite{instructGPT2022}, which aims to align language models with user intent and thus create useful assistants and chatbots. These days, many models release an instruction fine-tuned version along with the pre-trained model.

Fine-tuning large models with billions of parameters can be time and resource-intensive. Several Parameter-Efficient Fine-Tuning (PEFT) methods have been proposed that allow the training of fewer parameters while retaining the knowledge of the base foundational model. The method used in this work is a version of Low-Rank Adaptation (LoRA)~\cite{hu2022lora}. LoRA is a supervised fine-tuning method that introduces low-rank matrices that adapt the model's weights during training. The result of training with LoRA is the trained adapters, which can then be merged into the base model.

A further step towards alignment of the model with human goals is Reinforcement Learning from Human Feedback (RLHF), which requires the creation of a reward model that can evaluate the model's responses. Once trained, the reward model can be used with any reinforcement learning algorithm to update the base model to align with human preferences. RLHF requires datasets that contain human feedback and ratings of language model responses.


When faced with a complex task, humans tend to deconstruct it into simpler subtasks and solve them individually. Pre-trained language models, especially earlier versions such as GPT-3, were shown to have limited abilities when it comes to logical reasoning and planning. However, providing one or more examples as input can improve the model's ability to answer questions requiring reasoning~\cite{brown_language_2020}. The idea of guiding or teaching the model about the expected behavior during inference time using prompts is called In-context Learning (ICL). In contrast to fine-tuning, ICL does not require any training and places the task of teaching the model to the user that supplies the prompts.  

Leveraging a few demonstrative examples to guide a language model's behavior is known as \textit{k-shot} example learning, where $k$ represents the number of examples provided. The core idea is that by exposing the model to a handful of high-quality examples illustrating the desired reasoning process and expected outputs, the model can learn to generalize and apply that knowledge to solve novel problems.

One popular technique explored to improve k-shot example learning is Chain of Thought (CoT)~\cite{wei_chain-thought_2022}. The CoT approach provides the model with example inputs and outputs and explicitly demonstrates the step-by-step reasoning used to arrive at the final answer. The goal of explicitly showcasing the thought process is to help the model better understand the underlying logic and gain insights into the reasoning required to solve the given problem.

\section{Related Work}
\label{sec:related_work}
\subsection{Planning and Exploration}

Pre-trained large language models (LLMs) have shown some reasoning capabilities, using prompting techniques and ICL. However, they often face challenges in long-term planning, leading to occasional hallucinations and ineffective or irrelevant actions. Several frameworks have been developed to address these limitations that employ multi-stage prompting to enhance LLM agents' planning abilities by integrating reasoning and self-reflection. Notable examples include ReAct~\cite{yao2023react}, Reflexion~\cite{shinn_reflexion_2023}, Describe, Explain, Plan and Select (DEPS)\cite{wang_describe_2023}, and Reasoning via Planning (RAP)\cite{hao2023reasoning}.

ReAct~\cite{yao2023react} combines reasoning with action. Reflexion~\cite{shinn_reflexion_2023} advances this concept by introducing a sequential decision-making framework that includes self-reflection and evaluation, assessing the quality of actions and trajectories within an episode. This framework utilizes short-term memory to track actions during an episode and long-term memory to inform future decisions, allowing agents to learn from past experiences.

The RAP framework~\cite{hao2023reasoning} takes a different approach and uses an LLM as a world model to simulate actions and evaluate outcomes. RAP uses Monte Carlo Tree Search (MCTS) to explore various reasoning paths, with the LLM incrementally building a reasoning tree that considers the most promising steps. By leveraging the world model to predict potential outcomes, RAP helps the LLM refine its reasoning process through rewards derived from these outcomes. This approach allows the agent to simulate and anticipate the consequences of different actions, improving its planning and decision-making abilities.

In addition to planning tasks, LLM-based agents have been successful in exploration, as demonstrated by Du et al.\cite{du_guiding_2023} and Wang et al.\cite{wang2024voyager}. Voyager used an automatic curriculum, a skill library, and an iterative prompting mechanism for open-ended exploration in the Minecraft game environment. Similarly, Du et al. proposed using pre-trained LLMs to provide "intrinsic motivation" that guides the exploration and goal setting of the agent in the Crafter\cite{hafner2022benchmarking} and Housekeep\cite{kant_housekeep_2022} environments.

Other works have also explored using LLMs for planning tasks, such as Spring~\cite{spring2023wu}, which uses an LLM to "study" a paper describing the Crafter game environment. Using the summarized knowledge from the paper, it employs a guided Q\&A approach with the LLM to select the best action.

Earlier research has explored using LLMs for classical planning tasks~\cite{valmeekam2022large,silver2022pddl}, particularly within Planning Domain Definition Language (PDDL) domains. These studies highlighted the challenges of generating plans across diverse domains. Silver et al.~\cite{Silver2024generalized} demonstrated improved performance by using LLMs to generate plans as Python programs, refining them through debugging. However, classical planning approaches are limited to fully observable environments, which do not apply to cybersecurity domains' complex and dynamic settings.

\subsection{Red Teaming}
\label{pentesting}
The integration of Large Language Models (LLMs) in cybersecurity, particularly in automating penetration testing and red teaming, is a rapidly developing area of research. Previous studies underscore the diverse capabilities of LLMs within this realm and illustrate their potential and limitations.

Happe et al. (2023) propose a structured approach to penetration testing by categorizing tasks into higher and lower levels: higher-level tasks focus on planning and orchestration, while lower-level tasks involve executing specific attack tools. However, the practical application of their findings remains constrained, as the attacks were evaluated in a limited set of scenarios \cite{happe2023getting}.

Moskal et al. (2023) further explore the capabilities of LLMs by creating a controlled environment with a single target and employing a heuristic planner to assess LLMs’ performance in basic penetration testing tasks \cite{moskal2023llms}. Their work provides an initial framework for evaluating how LLMs might handle specific tasks in a simplified setting.

More recently, the PentestGPT framework\cite{deng2023pentestgpt} seeks to automate penetration testing more comprehensively, with evaluations conducted across various machines on platforms like HacktheBox. This approach indicates a more practical application of language models in varied environments. However, it is also focused more on the low-level tasks.

Raman et al. (2024) investigate the performance of commercial LLMs in the context of Certified Ethical Hacking (CEH) certification questions\cite{raman2024ceh}. Similarly, Tann et al. (2023) analyze the application of LLMs in Cisco Networking certifications and Capture The Flag (CTF) challenges~\cite{tann2023using}. While important, these evaluations do not necessarily translate to practical automation capabilities in realistic penetration testing situations; however, they provide insight into the models' understanding of security and networking concepts.

Overall, while findings related to certification exams and controlled environments contribute valuable insights into LLMs' knowledge regarding security concepts, they do not fully address the complexities of automating penetration testing in dynamic and diverse setups. Thus, continued exploration in realistic scenarios remains imperative for realizing the full potential of LLMs in cybersecurity.

\section{The NetSecGame Environment}
\label{sec:netsec_env}
NetSecGame is a network security simulation environment that follows a reinforcement learning (RL) architecture~\cite{rigaki_out_2024,Garcia_NetSecGame_a_RL}. The NetSecGame is designed to allow agents to play in a network by submitting high-level actions and providing back an observation of the current state of the environment for that agent. This allows RL agents to play, but it is not limited to agents. Humans can also play using an interactive interface with or without collaboration with an LLM-based assistant. Allowing humans to play is critical for evaluating attacking and defending strategies.


\paragraph{Network Configuration} It is possible to configure NetSecGame in different \textit{scenarios}. Each scenario has any number of host computers, services on each host, data on each host, routers, and how they are connected. It is also possible to define a host on the internet.  Each scenario also has a specific goal (can be fixed or random), starting position of the attacker agent (can be random), probabilities of detection of each action, and probabilities of success of each action.

\paragraph{Game State}
The objects defined in the environment are \textit{Networks}, \textit{IPs}, \textit{Services}, and \textit{Data}. The state is defined as a list of: \textit{known networks}, \textit{known hosts}, \textit{controlled hosts}, \textit{known services}, and \textit{known data}. \textit{Known networks} are the networks the agent knows about, the \textit{known hosts} are the hosts known to the agent because it scanned for them, the \textit{controlled hosts} are the hosts that the agent exploited successfully, and now controls, the \textit{known services} are the services (ports) known for the agent after scanning them, and the \textit{known data} are the data known to the agent because it scans for it inside the host.

An example of the state representation as given to the agents: 

\begin{tcolorbox}[colback=gray!5!white,colframe=gray!75!black]
\begin{small}
\begin{verbatim}
State<nets:{172.18.0.0/24, 172.18.2.0/24, 172.18.1.0/24}; 
known_hosts:{172.18.0.72, 172.18.0.55, 172.18.0.62, 172.18.0.9,
    172.18.1.25, 172.18.0.156, 43.82.9.214, 172.18.0.76};   
controlled_hosts:{172.18.0.72, 172.18.0.55, 43.82.9.214}; 
konwn_services:{172.18.0.72: 
    {Service(name='openssh', type='passive', version='8.1.0', 
    is_local=False)},
172.18.0.55:
    {Service(name='postgresql', type='passive', version='14.3', 
    is_local=False)
    }; 
known_data:{
    172.18.0.55: { Data(owner='User1', id='DatabaseData')}}>
\end{verbatim}
\end{small}
\end{tcolorbox}

\paragraph{Goals} The goal of each scenario is flexible and can be defined as a specific state, e.g., the presence of data in a specific host or a specific host that needs to be controlled by the agent. Randomization of the goal per episode is also an option. For example, the data that needs to be present in a specific host can change in each episode. Such approach can be used to evaluate generalization capabilities of the agents.

\paragraph{Actions} The action space has five \textit{action types}, each of which can receive parameters. The action types are: \textit{ScanNetwork}, \textit{FindServices}, \textit{FindData}, \textit{ExploitService}, and \textit{ExfiltrateData}. \texttt{ScanNetwork} receives as parameters the destination network; \texttt{FindServices} requires the destination host; FindData requires the destination host; \texttt{ExploitService} receives destination service and host; \texttt{ExfiltrateData} receives the source host, target host to exfiltrate, and target data to exfiltrate. Due to the action parameterization, the size of the action space depends on the number of objects defined in a specific scenario. The success probability for all actions was set to $1.0$ in all scenarios selected for this work.


\paragraph{Rewards} The rewards of each action are sparse. Each action that does not lead to the goal state or detection is rewarded with $-1$; if the goal state is reached, the agent gets an additional reward of $100$. A detection by the defender agent terminates the episode and has a reward of $-50$. However, these rewards sent by the environment can be ignored by the agents, who can define their internal reward functions if needed.

\paragraph{Stochastic Defender} In this version of NetSecEnv, the agent plays the role of an attacker. However, the environment allows for an omnipresent stochastic defender. The defender is aware of all actions and blocks the attacker agent if certain conditions are met:
Firstly, the defender uses a fixed-size time window to detect repeated actions. A threshold of maximum repeats is defined for each action type and upon reaching the threshold, the detection is triggered. Detection thresholds can be defined in the scenario configuration.  

Secondly, for Action types \textit{FindData}, \textit{ExploitService}, and \textit{ExfiltrateData}, there is an additional threshold for repeated actions \textbf{in the complete episode}. The episodic repeat detection mechanism uses both the action type and parameters for checking, which means the detection is triggered only if the same action is used multiple times by the agent.

Lastly, for action types \textit{ScanNetwok} and \textit{FindServices}, there is a check for consecutive actions. Inspired by port scanning detectors, this detection mechanism evaluates how many actions of type \textit{ScanNetwok} or \textit{FindServices} were played \textbf{in a sequence} and if the threshold is reached, the episode is terminated.

Each action has a different probability of detection, and all the parameters on the stochastic defender are configurable.

\section{LLM Agent Design}
\label{sec:agent_design}

\begin{figure*}
    \centering
    \includegraphics[width=0.9\linewidth]{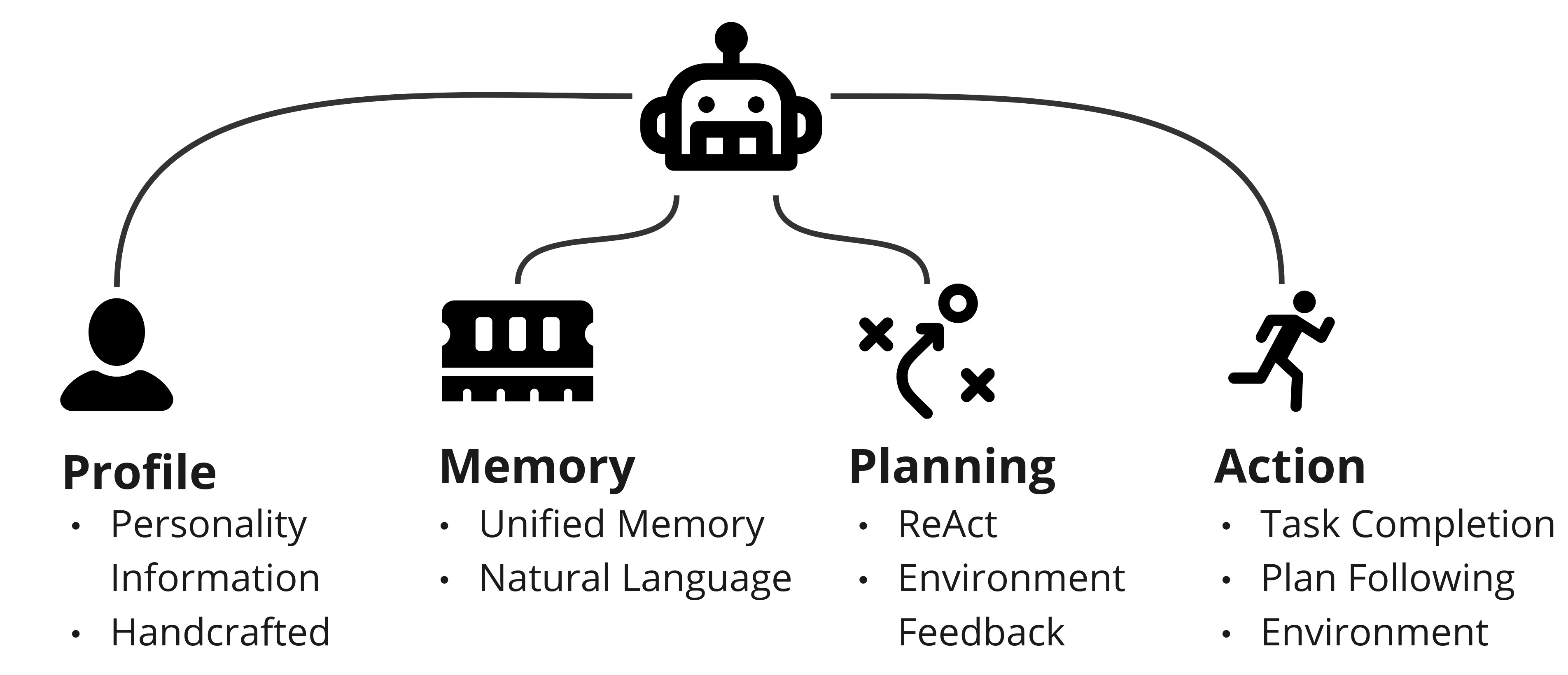}
    \caption{LLM Agent Components}
    \label{fig:agent_design}
\end{figure*}

The LLM agent used for all the experiments was introduced in~\cite{rigaki_out_2024}. As part of this work, we re-used the prompts and structure of the agent to directly compare the fine-tuned LLMs introduced in this work with the results of the commercial pre-trained models introduced in ~\cite{rigaki_out_2024}.  

Following the taxonomy proposed by~\cite{wang2024survey}, the LLM agent architecture comprises the following components: Profile, Memory, Planning, and Action (Figure~\ref{fig:agent_design}).

\paragraph{Profile} This component contains all the information provided to the models as \textit{personality information}. In our case, the personality of a penetration tester was \textit{handcrafted} in the prompt.

\paragraph{Memory} The memory component is a list of the previous actions that were taken by the agent and the current state of the environment: $\mathcal{M}=\{a_{t-k}, ..., a_{t-1}, s_t\}$. All the memory elements are presented as part of the prompt (\textit{unified memory}). The list of actions has a fixed size and contains the last $k$ actions the agent took. The agent is also provided with the current state of the environment, which is considered part of the agent's short-term memory. The agent does not use long-term memory between the episodes; therefore, each episode starts with an initial state and no action memory.

\paragraph{Planning} The agent uses the ReAct framework for planning and selecting the next actions. ReAct involves reasoning as the first step of the process and acting as the second step. Based on the current state of the environment $s_t$ the agent asks the LLM to analyze the objects in the state (IPs, Networks, Services, Data) and reason about the actions that can be taken for each object. In the second step, the agent uses the LLM response and the memory elements to decide and propose an appropriate action $a_t$. To produce a valid action, the agent is also provided with examples of actions (\textit{In-Context Learning}). The memory incorporates feedback as an \textit{intrinsic} reward. The agent evaluates each action as "helpful" or "not helpful" based on whether or not the state $s_{t+1}$ is different than $s_t$. Since the environment is deterministic and no action can remove objects from the state, an action is considered "helpful" if it allows the agent to discover more objects and reach a new state.


\section{Supervised Fine-tuning Methodology}
\label{sec:methodology}

\begin{figure*}
    \centering
    \includegraphics[width=0.99\linewidth]{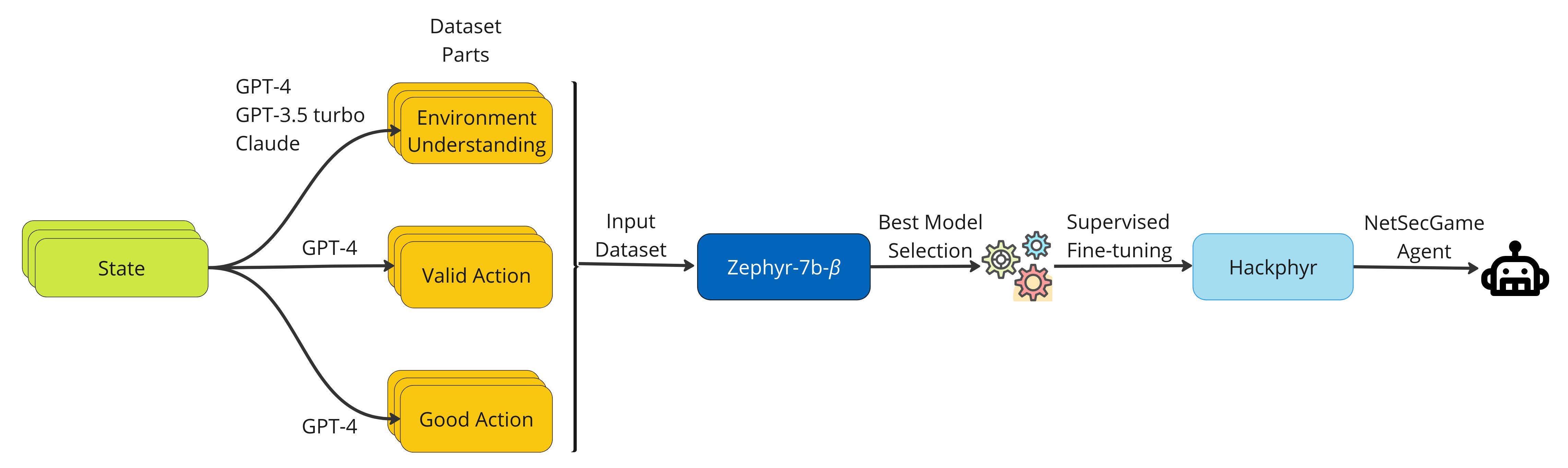}
    \caption{Supervised Fine-tuning Methodology}
    \label{fig:methodology}
\end{figure*}

This study was conducted through several steps (Figure~\ref{fig:methodology}). The first step of the process was to identify the weaknesses of the base models and create a dataset for supervised fine-tuning. The details of the dataset creation are presented in Section~\ref{sec:dataset-creation}. The base model used was Zephyr-7b-$\beta$~\cite{tunstall_zephyr_2023} which is an instruction-tuned model based on Mistral-7B-v0.1~\cite{jiang2023mistral}. The reasons the Zephyr model was selected were multiple: firstly, an already instruction-tuned model did not require time and effort to perform the instruction tuning. Secondly, the model had a good ability to produce valid JSON strings, and thirdly, in our initial experiments it showed that it has enough background knowledge of networking and security. Finally, we decided to select an open-source model with seven billion parameters so that it would be possible to fine-tune it using one GPU, which allows reproducibility of our results with minimal cost. The methodology followed in this paper is general enough and can be applied to other models of similar strength. 

Supervised fine-tuning is a training process that involves many hyperparameters. The method followed in this work is based on the Quantized LoRA (QLoRa)~\cite{qlora2023}. QLoRA allows a quantized version of the base model to be used during training requiring less memory in the GPU card. The rank of the adapters $r$ and the weight $\alpha$ are the two main hyperparameters used in LoRA, and the details about their selection are presented in Section~\ref{sec:best-model-selection}. For the fine-tuning process, we used the HuggingFace alignment-handbook library and scripts~\cite{alignment_handbook2023}.

After the hyperparameter tuning, we executed a series of experiments to evaluate the performance of our fine-tuned model. These experiments compared Hackphyr against established baselines across three distinct scenarios (Section~\ref{sec:scenarios}). All the supervised fine-tuning and the scenario experiments were executed in a single V100 Nvidia GPU card with 32GB of NVRAM.

\subsection{Dataset Creation}
\label{sec:dataset-creation}

Supervised fine-tuning requires high-quality data that help the language model perform well in a narrow task. The strategy for creating the dataset addressed issues we observed using the smaller models in the NetSecGame environment. The process was automated initially by asking stronger LLMs to play the role of a teacher who generates questions and answers to teach their students about the NetSecGame environment. After the automated generation of the questions and answers, we performed a human evaluation to ensure that the answers were of good quality, edited the incorrect ones, and discarded duplicates.

The dataset comprises 1641 questions and answers generated as three separate parts (Figure~\ref{fig:methodology}). Each part has a different focus and aims to address a specific problem observed while we were testing various open-source LLMs. The complete dataset was published on HuggingFace and is accessible for further research and experimentation~\cite{netsecdataset}.

\subsubsection{Part I - Environment Understanding}
The first part of the dataset contains questions and answers that test the model's ability to understand the current status of the environment and the provided rules. The dataset was generated automatically using three different LLMs: GPT-4, GPT-3.5-turbo from OpenAI, and Claude from Anthropic. The total number of questions and answers generated was $1080$.

Given a set of 18 states collected by previous game runs, the models were asked to generate 20 questions that test a student's ability to comprehend the game environment and the rules: 

\begin{tcolorbox}[colback=green!5!white,colframe=green!75!black]
\begin{small}
\begin{verbatim}
Provide 20 questions and answers that test the students
knowledge regarding the specific status and rules. 
\end{verbatim}
\end{small}
\end{tcolorbox}

An example of a generated question and answer can be found in Appendix~\ref{subsec:appendix_dataset_partI}.

\subsubsection{Part II - Generating Valid Actions}


The second part of the dataset contains questions that test the model's ability to generate \textbf{valid} actions, both in terms of syntax (JSON format) and in terms of semantics (validity in the specific state). The final number of questions and answers in the second part of the dataset is $450$. The following prompt generated this part of the dataset from 18 distinct states:

\begin{tcolorbox}[colback=green!5!white,colframe=green!75!black]
\begin{small}
\begin{verbatim}
Provide 20 questions and answers that test the student's
ability to generate valid and well-formatted actions in the
current status.
\end{verbatim}
\end{small}
\end{tcolorbox}
An example of a generated question and answer can be found in Appendix~\ref{subsec:appendix_dataset_partII}.

\subsubsection{Part III - Generating Good Actions}
This dataset part aims to teach the fine-tuned models how to make correct decisions given a specific environment state. It comprises 113 questions and answers generated from previous game runs using GPT-4 in the small scenario~\cite{rigaki_out_2024}. 

The state-action pairs were selected in a way where only actions that led to a new state of the environment were used. This choice is because actions that do not lead to a new state, i.e., the discovery of a new element, are either repetitions or not useful. 

An example of a generated question and answer can be found in Appendix~\ref{subsec:appendix_dataset_partIII}.

\subsection{Hyper-parameter Tuning}
\label{sec:best-model-selection}
Instead of the common machine learning evaluation approach of splitting between train and test sets, the complete dataset was used for training following the SFT methodology described in section~\ref{sec:background}. Since random IPs and random data exfiltration goals were set for each episode, the agent is evaluated under different conditions than the ones in the training dataset. 

Then, the resulting model was tested on the small scenario~\ref{sec:scenarios} for 150 episodes using the LLM-based agent as described in section~\ref{sec:agent_design} with a memory of the last ten actions. 

A hyperparameter grid search was conducted for selecting the LoRA $r$ and $\alpha$ hyperparameters. The following parameter values were used during the grid search:

\begin{itemize}
    \item LoRA r: 4, 8, 16, and 32
    \item LoRA $\alpha$: 4, 8 16 and 32
\end{itemize}

The rest of the hyper-parameters were set to default values according to the HuggingFace alignment handbook~\cite{alignment_handbook2023}. 

\begin{figure}[t]
    \centering
    \includegraphics[width=1\linewidth]{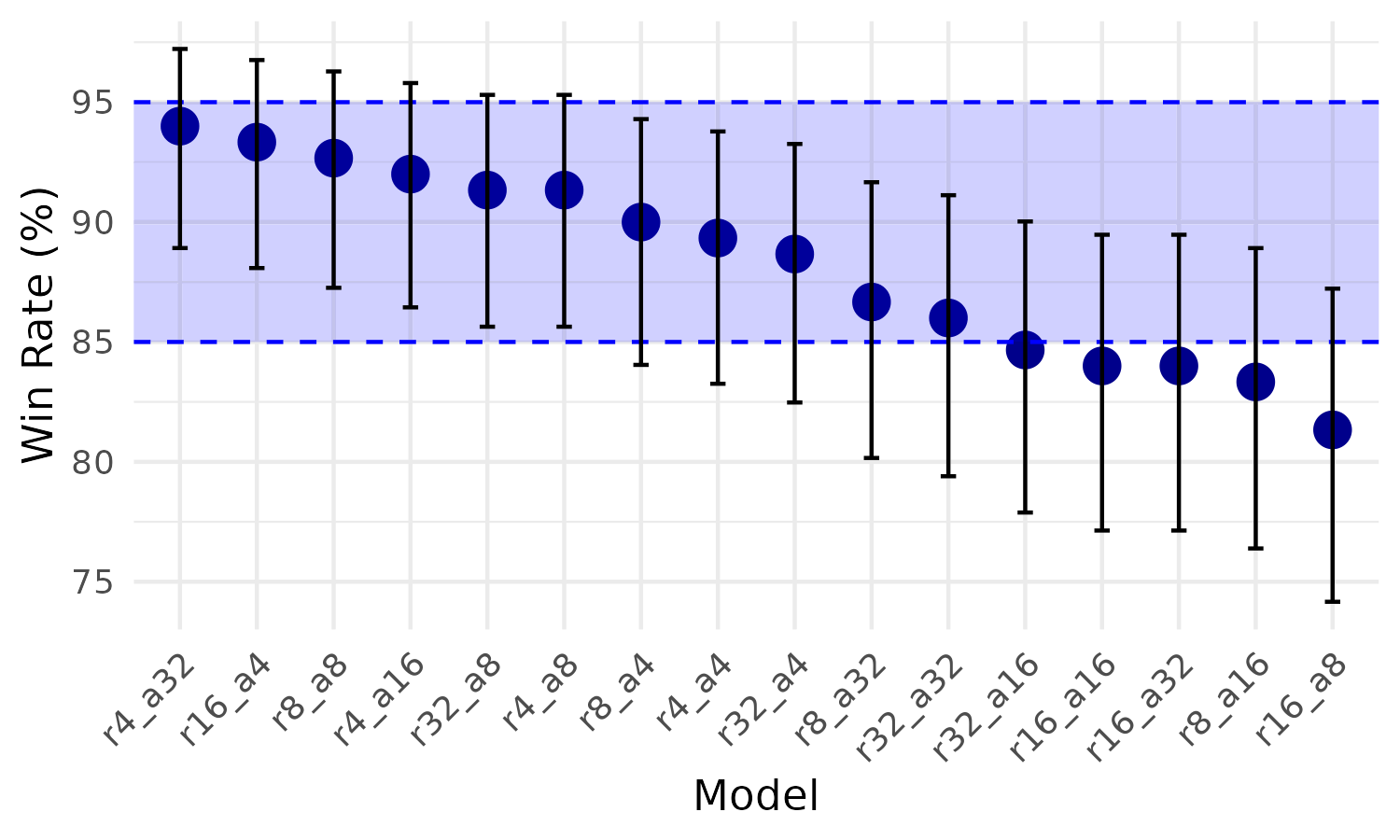}
    \caption{Win rate(\%) confidence intervals by model. The blue area indicates the desired effect size.}
    \label{fig:confidence-intervals}
\end{figure}

The performance of each agent in terms of the win rate, along with the confidence intervals (CI) over the 150 episodes, is shown in Figure~\ref{fig:confidence-intervals}. The number of 150 episodes was chosen based on a power analysis, with significance level $\alpha=0.05$ and power $1-\beta=0.8$ aiming for an effect size between 85\% and 95\%. The purpose of using confidence intervals is to focus on those models whose intervals do not overlap. 

From figure~\ref{fig:confidence-intervals}, we can observe that only the last five hyperparameter combinations showed significant differences between the analyzed effect sizes. The remaining hyperparameters do not provide strong enough evidence to conclude a significant difference between the win rate values at the given confidence level. However, the considerable overlap of the CIs suggests that the sample data do not provide strong enough evidence to conclude a significant difference between the win rates at the given confidence level. Therefore, we selected the agent that used a model with LoRa parameters set to $r4\_a32$, as this configuration consistently demonstrated the highest average win rate.

The final values of the most important hyper-parameters are provided in Table~\ref{tab:hyper-params}. 

\begin{table}
  \caption{Final hyper-parameter values}
  \label{tab:hyper-params}
  \centering
  \begin{tabular}{lc}
    \toprule
    Parameter & Value \\
    \midrule
    LoRA r & 4 \\
    LoRA $\alpha$ & 32 \\
    training epochs & 2 \\
    gradient accumulation steps & 2 \\
    learning\_rate & 2.0e-04 \\
    lr scheduler type & cosine \\
    max seq length & 2048 \\
    number of GPUs & 1 \\
    training batch size & 1 \\
    \bottomrule
    \end{tabular} 
\end{table}

\section{Experiment Design}
\label{sec:experiment_design}

For evaluating the agents, we have selected three scenarios on top of the NetSecGame environment described in Section~\ref{sec:netsec_env}. The goal in all the scenarios was to exfiltrate specific data to the C\&C server. Agents must perform a sequence of actions to discover data on a computer in the network and then exfiltrate it.

\subsection{Scenario Descriptions}
\label{sec:scenarios}

\begin{figure}[t]
    \centering
    \includegraphics[width=0.60\linewidth]{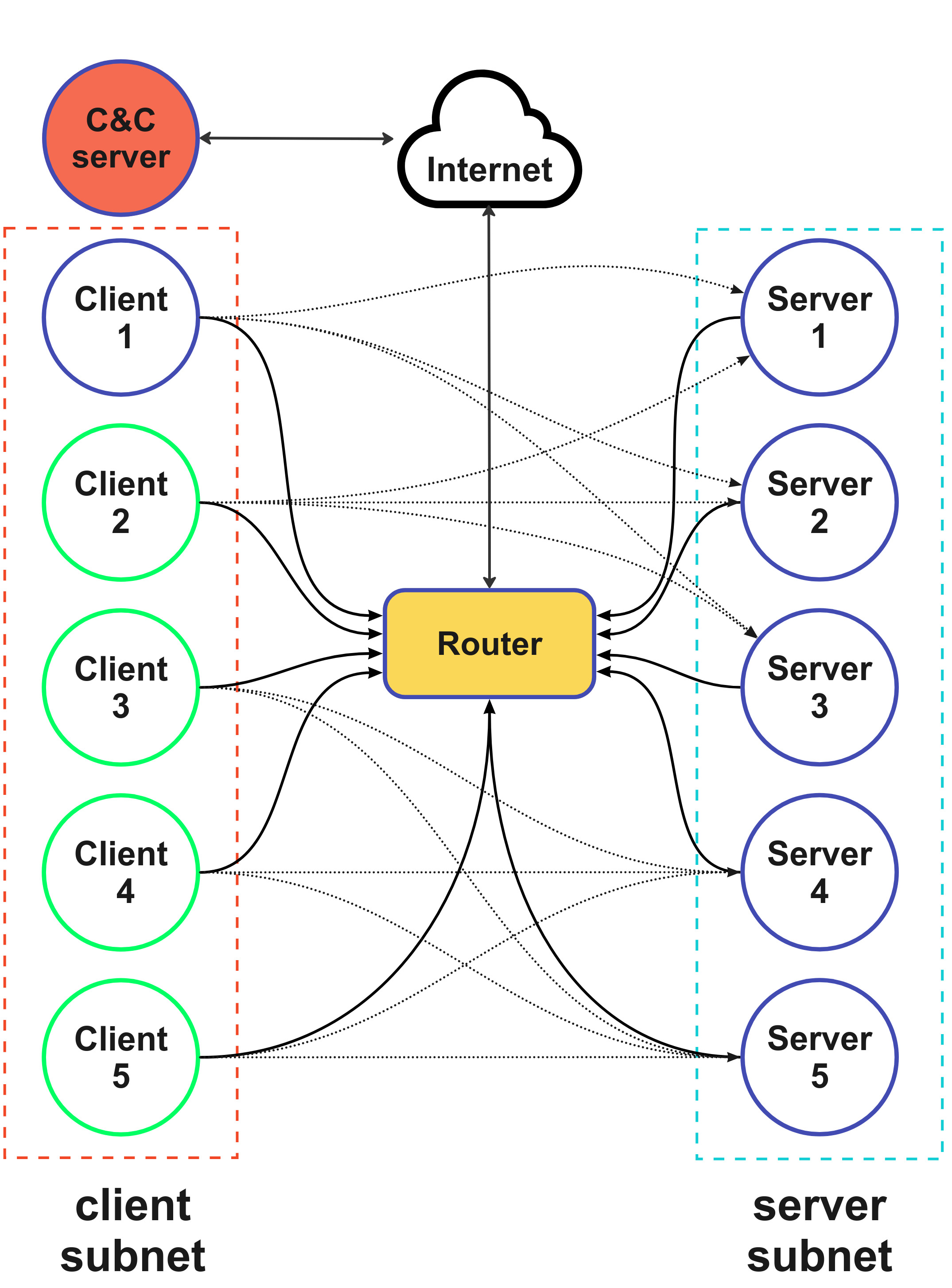}
    \caption{Small and Full network scenarios~\cite{rigaki_out_2024}. The small scenario has only one client in the client subnet, while the full scenario has all five clients.}
    \label{fig:base_scenario}
\end{figure}

The first two data exfiltration scenarios (denoted as \textit{small} and \textit{full} scenarios) are depicted in Figure~\ref{fig:base_scenario}. These two scenarios share the same network topology: two subnets, one with clients and one with servers, connected by a router. The small scenario differs from the full scenario only in that there is a single client on the client subnet. 

The attacker starts the scenario in one of the client machines, playing the role of an attacker who has gained a foothold in a network. The attacker also controls a machine on the internet, where the command and control (C\&C) server is hosted. 


Both scenarios require a minimum of five steps to exfiltrate the data. Still, the solution is far from trivial, given that the IP addresses and goals change in every episode, and the latter presents a challenge to both LLM and traditional RL agents. A more detailed description of both scenarios is presented in~\cite{rigaki_out_2024}.

\begin{figure}
    \centering
    \includegraphics[width=0.70\linewidth]{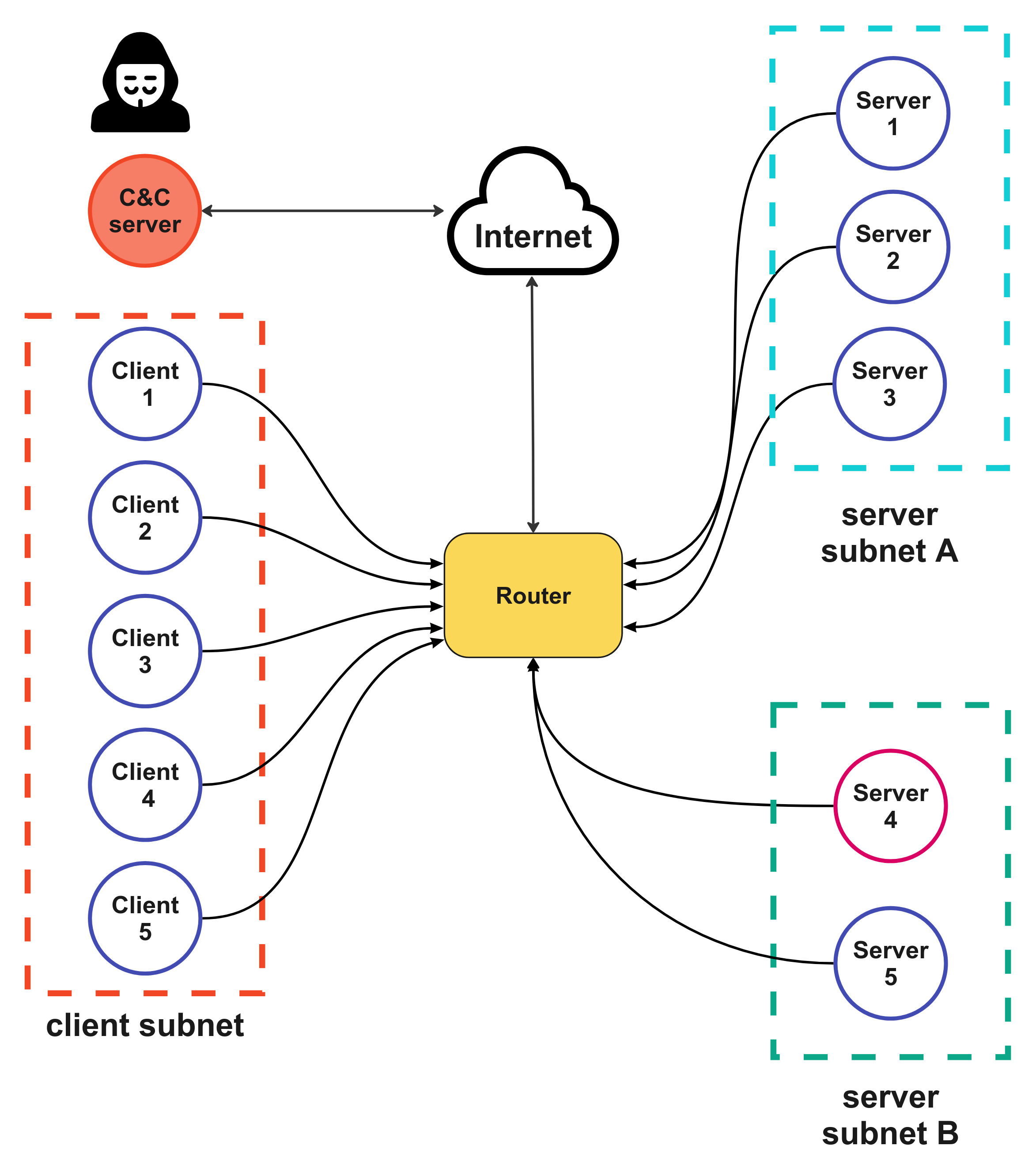}
    \caption{Three Subnets Scenario Topology. The clients can only access subnet A directly. The goal is to exfiltrate data from subnet B by first gaining a foothold to subnet A.}
    \label{fig:3_nets_scenario}
\end{figure}

The third scenario (denoted as the three-networks scenario) consists of three different subnetworks inside a fictional small-medium enterprise (Figure~\ref{fig:3_nets_scenario}). The client subnet contains the client hosts (PCs, etc.). The first server subnet (subnet A) contains some servers that are immediately accessible by the clients, and the second server subnet (subnet B) contains two servers that are only accessible by subnet A. Firewall rules prevent the clients from accessing the servers in subnet B directly. The attacker aims to exfiltrate the data from one of the servers in the subnet B to the C\&C server on the internet. The attacker begins with a foothold in the client subnet and must first gain control of one of the servers in subnet A. Then they have to perform a network scan of network B to discover and exploit the servers and gain control. Once the correct server is controlled and the data are discovered, they can be exfiltrated to the C\&C server. 

The three-networks scenario requires at least three more steps to solve than the previous two. This scenario is more demanding, and it is used to test the limits of the LLMs.  
As in the previous two scenarios, the IP addresses were randomized in each episode, and the scenarios were tested with and without the stochastic defender with thresholds; however, in this scenario, the goal is not randomized. 

A summary of the main differences between the three scenarios is presented in Table~\ref{tab:scenarios}

\begin{table*}[t]
\centering
\begin{tabular} {p{2.0cm}p{2.9cm}p{2.9cm}p{2.9cm}}
\toprule
\textbf{Scenario} & \textbf{Small} & \textbf{Full} & \textbf{Three-Networks} \\
\midrule
\textbf{Network Topology} & \footnotesize One client and one server subnet, router & \footnotesize One client and one server subnet, router & \footnotesize One client and two server subnets (A \& B), client have direct access only to subnet A, router \\ 
\textbf{Config} & \footnotesize Single client and multiple servers & \footnotesize Multiple clients and multiple servers & \footnotesize Multiple clients and multiple servers \\ 
 
\textbf{Episode Setup } & \footnotesize Randomized IPs and goal   & \footnotesize Randomized IPs and goal & \footnotesize Randomized IPs, fixed goal on subnet B \\ 

\textbf{Steps} & \footnotesize  Five & \footnotesize  Five & \footnotesize Eight \\
\midrule
\midrule
\textbf{Purpose} & \footnotesize Fine-tuning and hyper-parameter tuning & \footnotesize Test performance under more complex conditions & \footnotesize Test adaptability to different network configurations \\ 

\bottomrule
\end{tabular}
\caption{Comparison of Data Exfiltration Scenarios}
\label{tab:scenarios}
\end{table*}

\subsection{Evaluation Procedure}

Each scenario was designed with a specific purpose in mind for facilitating the evaluation of the agents' performance across different level of complexity levels and  conditions (See Table~\ref{tab:scenarios}).

In particular, the small scenario was mainly used for fine-tuning and hyper-parameter selection, as similar examples were used during the model's training. We tested this scenario with and without a defender to establish a performance baseline.

The full scenario is more complex and involves more clients that can be scanned for services and finding the data to exfiltrate. Since the full scenario has more clients and was not seen during training, we used it to analyze the model's performance degradation under slightly more complex conditions.

Finally, the three networks scenario is significantly more complex, involving additional networks and more steps required for data exfiltration. This scenario introduced a different network topology not encountered during training, and it helped analyze the model for potential overfitting. This ensures that the fine-tuned models can adapt to diverse and demanding network configurations without retraining.

The language models used for testing the scenarios were GPT-4, GPT-3.5-turbo, and Zephyr-7b-$\beta$, along with Hackphyr. Similar to Section~\ref{sec:best-model-selection}, we evaluated the performance of each model over 150 episodes with 100 maximum steps per episode. Each episode was run independently and IPs and goals were generated with a different seed.

In addition, we conducted some baseline comparisons using a random agent and a tabular Q-learning agent~\cite{watkins_q-learning_1992} similar to~\cite{rigaki_out_2024}. For the Q-learning agent, it was not possible to randomize the IP per episode because the Q-table was fixed. However, the goal was randomized in each episode. The Q-agent was trained for 50,000 episodes, and the evaluation was performed on the trained agent.
The random agent was run for 2,000 episodes.

For all agents, we measured the percentage of wins (win\_rate), the average steps, and the average returns. A win is any episode where the agent reaches the goal state within the 100 maximum steps. 

\section{Results}
\label{sec:results}

The results of the experiments without and with the stochastic defender are presented in Tables~\ref{tab:performance_no_defender} and~\ref{tab:performance_with_defender}, respectively. The tables present the win rates and returns for the small, full, and three subnets scenarios.

For the small and full scenarios, the results were taken from~\cite{rigaki_out_2024}. For the three subnets scenario, all the LLM-based agents were run using exactly the same setup, including prompts, temperatures, and other relevant parameters. This also applied to the Q-learning agent, with settings such as the learning rate, epsilon, and the number of episodes being consistent across the experiments.

\subsection{Scenarios Without Defender} 


\begin{table*}[t]
\caption{Average returns and win rates of all agents across different scenarios without defender with 100 maximum steps per episode.}
\label{tab:performance_no_defender}
\centering
\begin{tabular}{lcccccc}
\toprule
& \multicolumn{2}{c}{Small} & \multicolumn{2}{c}{Full} & \multicolumn{2}{c}{3 Subnets } \\
\cmidrule(lr){2-3} \cmidrule(lr){4-5} \cmidrule(lr){6-7}
Agent & Win\% & Return & Win\% & Return & Win\% & Return \\
\midrule
Random & \cellcolor[rgb]{0.97,0.98,0.99} 40.99 & \cellcolor[rgb]{0.97,0.98,0.99} -43.43 & \cellcolor[rgb]{0.98,0.99,0.99} 31.16 & \cellcolor[rgb]{0.98,0.99,0.99} -57.49 & \cellcolor[rgb]{0.96,0.98,0.99} 4.86 & \cellcolor[rgb]{0.96,0.98,0.99} -93.28 \\
Q-learning & \cellcolor[rgb]{0.43,0.65,0.88} \color{white}67.41 & \cellcolor[rgb]{0.54,0.72,0.91} \color{white}47.55 & \cellcolor[rgb]{0.54,0.72,0.91} \color{white}58.74 & \cellcolor[rgb]{0.54,0.72,0.91} \color{white}48.00 & \cellcolor[rgb]{0.96,0.98,0.99} 0.00 & \cellcolor[rgb]{0.96,0.98,0.99} -100.00 \\
GPT-3.5-turbo & \cellcolor[rgb]{0.90,0.94,0.97} 50.00 & \cellcolor[rgb]{0.97,0.98,0.99} -16.13 & \cellcolor[rgb]{0.98,0.99,0.99} 30.00 & \cellcolor[rgb]{0.98,0.99,0.99} -51.67 & \cellcolor[rgb]{0.96,0.98,0.99} 10.00 & \cellcolor[rgb]{0.96,0.98,0.99} -87.50 \\
GPT-4 & \cellcolor[rgb]{0.26,0.49,0.69} \color{white}100.00 & \cellcolor[rgb]{0.31,0.54,0.74} \color{white}83.10 & \cellcolor[rgb]{0.26,0.49,0.69} \color{white}100.00 & \cellcolor[rgb]{0.34,0.57,0.77} \color{white}77.13 & \cellcolor[rgb]{0.31,0.54,0.74} \color{white}82.35 & \cellcolor[rgb]{0.85,0.91,0.96} 18.78 \\
Zephyr-7b-$\beta$ (base) & \cellcolor[rgb]{0.98,0.99,0.99} 30.46 & \cellcolor[rgb]{0.98,0.99,0.99} -51.80 & \cellcolor[rgb]{0.96,0.98,0.99} 23.65 & \cellcolor[rgb]{0.98,0.99,0.99} -51.77 & \cellcolor[rgb]{0.96,0.98,0.99} 1.75 & \cellcolor[rgb]{0.96,0.98,0.99} -98.80 \\
Hackphyr (ours) & \cellcolor[rgb]{0.28,0.51,0.71} \color{white}94.00 & \cellcolor[rgb]{0.34,0.57,0.77} \color{white}72.83 & \cellcolor[rgb]{0.31,0.54,0.74} \color{white}89.10 & \cellcolor[rgb]{0.43,0.65,0.88} \color{white}61.03 & \cellcolor[rgb]{0.54,0.72,0.91} 50.34 & \cellcolor[rgb]{0.97,0.98,0.99} -14.26 \\
\bottomrule
\end{tabular}
\end{table*}

Regarding win rate and average reward return, GPT-4 agents consistently outperform all the LLM agents in all the scenarios without defenders(Small, Full, and Three Subnets). On the other hand, the Hackphyr agents demonstrate strong performance, always ranking second to GPT-4 and significantly better when compared with the base Zephyr agent. 

In the small scenario, the performance of Hackphyr agents is close to GPT-4 (94\% win rate. 72.8 average return). These are expected results since the dataset used for fine-tuning Zephyr contained examples from actions taken by GPT-4-based agents on the small scenario. 

The performance in the full scenario is more valuable in terms of win rate and average return. This scenario was completely unseen for the Hackphyr agent during the fine-tuning process described in section~\ref{sec:methodology}. Despite the decrease compared to the small scenario, the Hackphyr agent showed a win rate of 89\% with an average return of 61. These values were exceptionally better than the Zephyr-based agent and outperformed GPT-3.5-turbo, a larger and more powerful model.  

Finally, the complexity of the three subnets scenario caused a decrease in the win rate performance of all language models. The Hackphyr dropped to a 50\% win rate and a -14.25 average return. The GPT-4-based agents observed a decrease in win rate of 20\% and a drop in average return of 18.78. The fact that a powerful model such as GPT-4 had a performance degradation shows that this is a harder scenario and that an improved agent design may be required.  

The performance of the two baseline agents, Random and Q-learning, highlights the limitations of simpler strategies in the given scenarios. The Random agent consistently underperforms across all scenarios, with win rates of 40.99\%, 31.16\%, and 4.86\% in the Small, Full, and Three Subnets scenarios, respectively, and corresponding negative returns, indicating poor decision-making that fails to generate positive outcomes. The Q-learning agent shows a notable improvement over the Random agent, particularly in the Small and Full scenarios, achieving win rates of 67.41\% and 58.74\%, respectively, with positive returns in both cases (47.55 and 48.00). However, in the more complex Three Subnets scenario, the Q-learning agent struggles significantly, with a 0\% win rate and a return of -100.00, reflecting the limitations of the tabular RL methods. The results underscore the limited effectiveness of these baseline methods, particularly in more challenging environments, compared to more advanced models like GPT-4 and Hackphyr.

In summary, the Hackphyr agents demonstrated remarkable performance, particularly considering their fine-tuning process and the complexity of the scenarios they faced. While GPT-4 consistently outperforms all other agents, Hackphyr stands out as a robust and adaptable model, achieving results that often come close to those of GPT-4, significantly better than the other LLM-based agents such as GPT-3.5-turbo.

\subsection{Scenarios With Defender}

\begin{table*}[t]
\caption{Average returns and win rates of all agents across different scenarios with defender with 100 maximum steps per episode.}
\label{tab:performance_with_defender}
\centering
\begin{tabular}{lccccccc}
\toprule
& \multicolumn{2}{c}{Small} & \multicolumn{2}{c}{Full} & \multicolumn{2}{c}{3 Subnets } \\
\cmidrule(lr){2-3} \cmidrule(lr){4-5} \cmidrule(lr){6-7}
Agent & Win\% & Return & Win\% & Return & Win\% & Return \\
\midrule
Random & \cellcolor[rgb]{0.96,0.98,0.99} 3.76 & \cellcolor[rgb]{0.96,0.98,0.99} -65.57 & \cellcolor[rgb]{0.96,0.98,0.99} 2.72 & \cellcolor[rgb]{0.96,0.98,0.99} -66.56 & \cellcolor[rgb]{0.96,0.98,0.99} 0.13 & \cellcolor[rgb]{0.96,0.98,0.99} -70.73 \\
Q-learning & \cellcolor[rgb]{0.31,0.54,0.74} \color{white}77.96 & \cellcolor[rgb]{0.43,0.65,0.88} \color{white}54.91 & \cellcolor[rgb]{0.34,0.57,0.77} \color{white}71.00 & \cellcolor[rgb]{0.54,0.72,0.91} \color{white}45.38 & \cellcolor[rgb]{0.96,0.98,0.99} 0.00 & \cellcolor[rgb]{0.96,0.98,0.99} -70.45 \\
GPT-3.5-turbo & \cellcolor[rgb]{0.93,0.96,0.98} 20.00 & \cellcolor[rgb]{0.93,0.96,0.98} -34.27 & \cellcolor[rgb]{0.93,0.96,0.98} 16.67 & \cellcolor[rgb]{0.93,0.96,0.98} -58.00 & \cellcolor[rgb]{0.93,0.96,0.98} 6.67 & \cellcolor[rgb]{0.93,0.96,0.98} -63.63 \\
GPT-4 & \cellcolor[rgb]{0.26,0.49,0.69} \color{white}83.33 & \cellcolor[rgb]{0.34,0.57,0.77} \color{white}58.83 & \cellcolor[rgb]{0.43,0.65,0.88} \color{white}53.33 & \cellcolor[rgb]{0.85,0.91,0.96} 8.80 & \cellcolor[rgb]{0.70,0.82,0.94} 36.36 & \cellcolor[rgb]{0.90,0.94,0.97} -21.69 \\
Zephyr-7b-$\beta$ (base) & \cellcolor[rgb]{0.96,0.98,0.99} 3.00 & \cellcolor[rgb]{0.96,0.98,0.99} -66.40 & \cellcolor[rgb]{0.96,0.98,0.99} 3.33 & \cellcolor[rgb]{0.96,0.98,0.99} -66.68 & \cellcolor[rgb]{0.96,0.98,0.99} 0.62 & \cellcolor[rgb]{0.96,0.98,0.99} -70.42 \\
Hackphyr (ours) & \cellcolor[rgb]{0.39,0.61,0.83} \color{white}59.77 & \cellcolor[rgb]{0.62,0.77,0.93} 33.00 & \cellcolor[rgb]{0.54,0.72,0.91} 44.00 & \cellcolor[rgb]{0.93,0.96,0.98} 3.56 & \cellcolor[rgb]{0.77,0.86,0.95} 23.33 & \cellcolor[rgb]{0.90,0.94,0.97} -37.67 \\
\bottomrule
\end{tabular}
\end{table*}

Table ~\ref{tab:performance_with_defender} shows the results with the stochastic defender. In all the cases, similar to the scenarios with a defender, the Hackphyr agents outperformed the agent based on Zephyr without fine-tuning.

For the small scenario, the Hackphyr agent showed a win rate of 59.77\% with a return value of 33. The GPT-4-based agent won 83 episodes with an average return value of 58. The agent using GPT-3.5-turbo was capable of winning in just 20\% of the episodes.

Similarly to the scenario without the stochastic defender, the performance decreased in the full and the three subnets scenarios. In the full scenario, the Hackphyr agent showed a win rate value of 44\% with an average return of 3.5. GPT-4-based agents performed slightly better, with a win rate of 53\% and an average return value of 8.8.

The three subnets scenario was difficult for all the agents. Despite the difficulties, the FT-Zephyr agent won in 23\% of the episodes, far beyond the results of a larger model such as GPT-3.5-turbo with only a 6\% win rate. GPT-4 only was able to win 36\% of the time. In all the cases, the average return was negative.   

When the stochastic defender was present, all the LLM-based agents' performance decreased. However, the Hackphyr agent performed relatively well, outperforming other LLM-based agents. Moreover, in the three subnets scenario, the Hackphyr agent outperformed all LLM-based agents (except for GPT-4) by a significant margin. This performance demonstrates that the model can succeed in a completely different scenario, indicating that it did not simply memorize data from the small scenario used for fine-tuning. This suggests a versatile and adaptable capability.

Regarding the baseline agents, it is worth mentioning that Q-learning displayed considerable effectiveness in scenarios with defenders, sometimes even exceeding the performance of both GPT-4 and Hackphyr. Specifically, it achieved a win rate of 77.96\% with an average return of 54.91 in the small scenario and a win rate of 71.00\% with an average return of 45.38.

These results could be attributed to the defender's presence, which helped the agent learn a policy that avoided repetitions penalized due to the detection. The resulting policy made the agent perform even better than in scenarios without a defender.  However, it faced significant challenges in the Three Subnets Scenario, highlighting potential limitations in more complex environments.

Finally, it must be noted that none of the LLM prompts had instructions to avoid the defender. Agents like the GPT-4-based sometimes followed a breadth-first approach, scanning hosts for services sequentially, which can trigger the stochastic defender.

\section{Behavioral Analysis}
\label{sec:behavioral_analysis}

We define a trajectory as an agent's sequence of actions during one episode. Formally, a trajectory $\tau$ can be expressed as:
\[
\tau = \{a_i\}_{i=1}^{T}
\]
where \( a_i \) denotes the action taken by the agent at time step \( i \), and \( T \) is the total number of time steps in the episode. Let \( \mathcal{T} \) represent a set of trajectories expressed as:
\[
\mathcal{T} = \{\tau_j\}_{j=1}^{N}
\]
where \( \tau_j \) denotes the \( j \)-th trajectory in the set, and \( N \) is the total number of trajectories.  

In this context, the goal was to analyze the set of trajectories \( \mathcal{T} \) to understand the rationality and correctness of the agent's behavior throughout the episodes. We performed a graph analysis to understand the transitions between actions taken by an agent that helped detect invalid or incorrect action transition patterns.

\begin{figure}[!t]
    \centering
    \begin{subfigure}[b]{0.45\textwidth}
        \centering
        \includegraphics[width=1.05\textwidth]{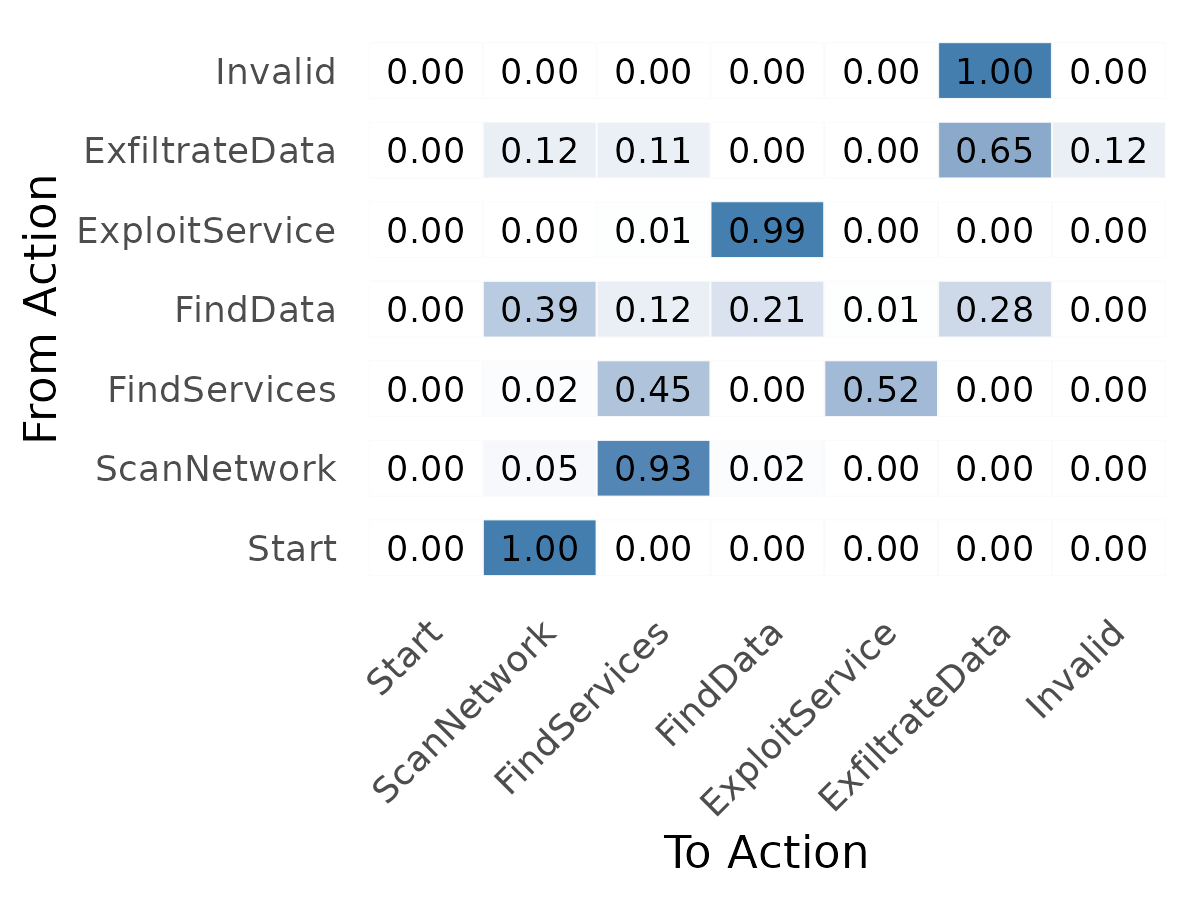}
        \caption{Transition Matrix}
        \label{fig:trans-matrix-gpt4}
    \end{subfigure}
    \begin{subfigure}[b]{0.45\textwidth}
        \centering
        \includegraphics[trim={5cm 3cm 3cm 2.5cm},width=0.85\textwidth]{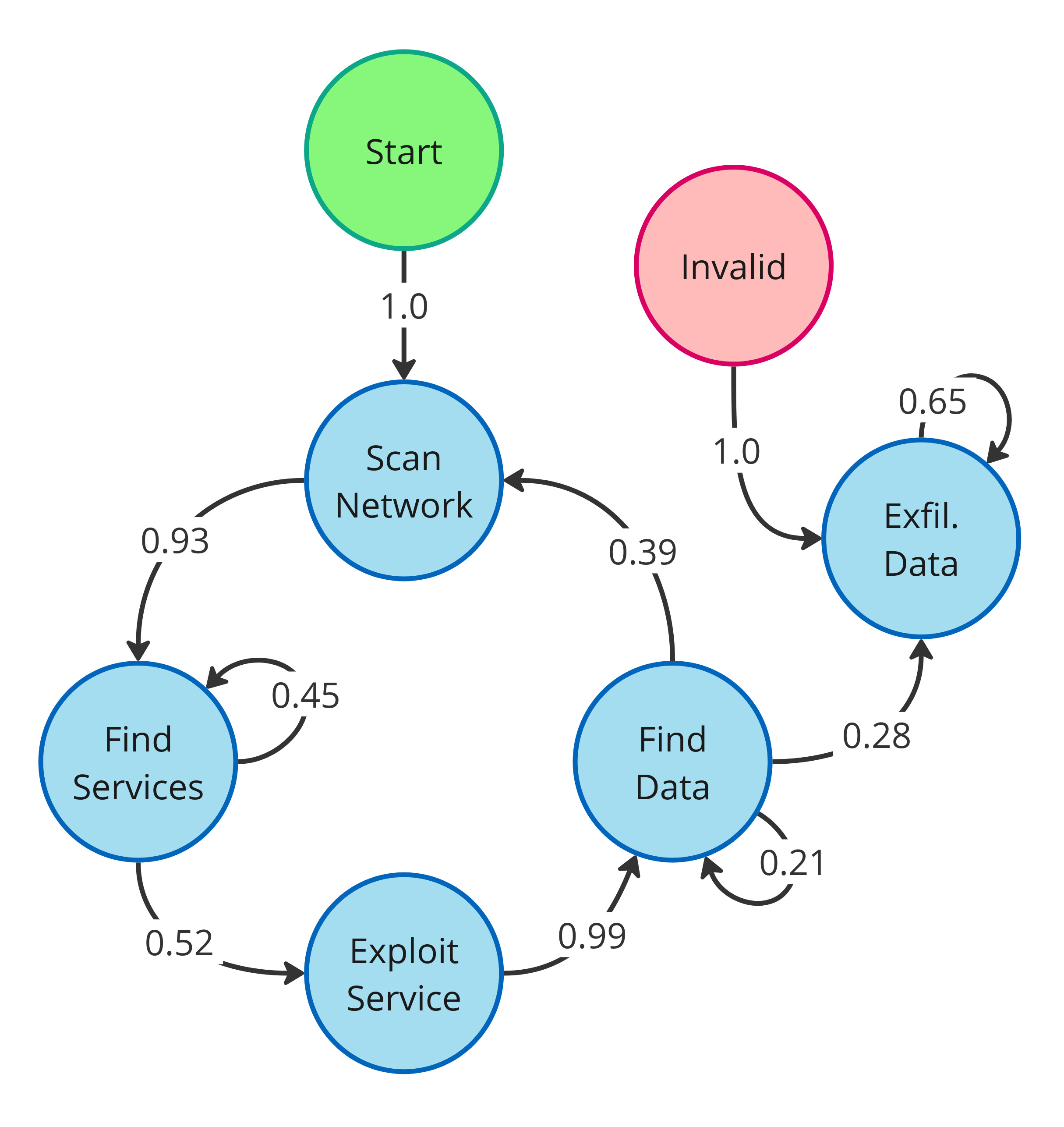}
        \caption{Key Action Transitions}
        \label{fig:key-action-gpt4}
    \end{subfigure}
    \caption{GPT-4 action transitions}
    \label{fig:behavior-gpt4}
\end{figure}

We considered GPT-4 the reference agent since it showed the best results in all the scenarios. Figure~\ref{fig:behavior-gpt4} illustrates the transition pathways the GPT-4 agent takes. The heatmap in Figure~\ref{fig:trans-matrix-gpt4} shows the probabilities of transitioning from one action to another, with higher probabilities highlighted in warmer colors. In Figure~\ref{fig:key-action-gpt4}, the key action transition graph visually represents the most relevant transitions (with probabilities greater or equal to 0.20) among the possible actions for GPT-4. Actions such as \texttt{Start} and \texttt{Invalid} are distinctly marked in green and red, respectively, to highlight that they are not actions an agent can choose from. 
In particular, an invalid action can be caused by semantic or syntactic errors. A semantic error occurs when the action taken is not allowed given the current environment state. On the other hand, a syntactic error is caused when the agent generates an invalid string representation of the action.

The transition matrix from Figure~\ref{fig:trans-matrix-gpt4} shows that the GPT-4-based agent is, in general, very confident when taking the next action. Most of the actions transition to the next with high probability. The only actions showing more evenly distributed transitions are the \texttt{FindData} and \texttt{FindServices}. 

When analyzing the key action transitions from the GPT-4-based agent (Figure~\ref{fig:key-action-gpt4}), the agent starts scanning the network with a probability of $1.0$. This action seems logical for potential targets or vulnerabilities. Then, transitioning from \texttt{ScanNetwork} to \texttt{FindServices} with a high probability of $0.93$ is expected, as identifying available services is a key step after scanning a network and identifying new hosts. 

The agent's transition from \texttt{FindServices} to \texttt{ExploitService} with probabilities of $0.5$ makes sense, as exploiting vulnerabilities in identified services is a typical next step after discovering those services. The self-loop in the \texttt{FindServices}  can be explained by the additional FindServices actions involving different IP addresses found during the ScanNetwork action.

The high probability ($0.99$) transition from \texttt{ExploitService} to \texttt{FindData} suggests that after exploiting a service, the primary goal is to find valuable data, which is logical in network intrusion scenarios.

Transitioning from \texttt{FindData} to \texttt{ExfiltrateData} with probabilities $0.28$ is logical, as extracting valuable data follows finding it. If the agent does not find any data, then a logical action is to start scanning again, and the GPT-4 agent goes to \texttt{ScanNetwork} with $0.39$ probability. The self-loop in the \texttt{FindData} mostly indicates that the action did not discover any data, and GPT-4 decided to search again.


The absence of outgoing transitions to \texttt{Invalid} actions suggests that the GPT-4-based agent has a low probability of generating \texttt{Invalid} actions from any other action. However, when an invalid action does occur, it is always associated with the \texttt{ExfiltrateData} action. This pattern could be due to the fact that \texttt{ExfiltrateData} is the most complex action, having the most parameters. Additionally, whenever an invalid action is generated, the agent's next action is consistently to \texttt{ExfiltrateData}, suggesting a strong correlation between the two.

The GPT-4-based agent shows a behavior pattern of systematic reconnaissance, service discovery, exploitation, data finding, and exfiltration that matches well-known attacker tactics and techniques like the ones in the MITRE ATT\&CK framework\footnote{https://attack.mitre.org/matrices/enterprise/}. 

\begin{figure}[!t]
    \centering
    \begin{subfigure}[b]{0.45\textwidth}
        \centering
        \includegraphics[width=1.1\textwidth]{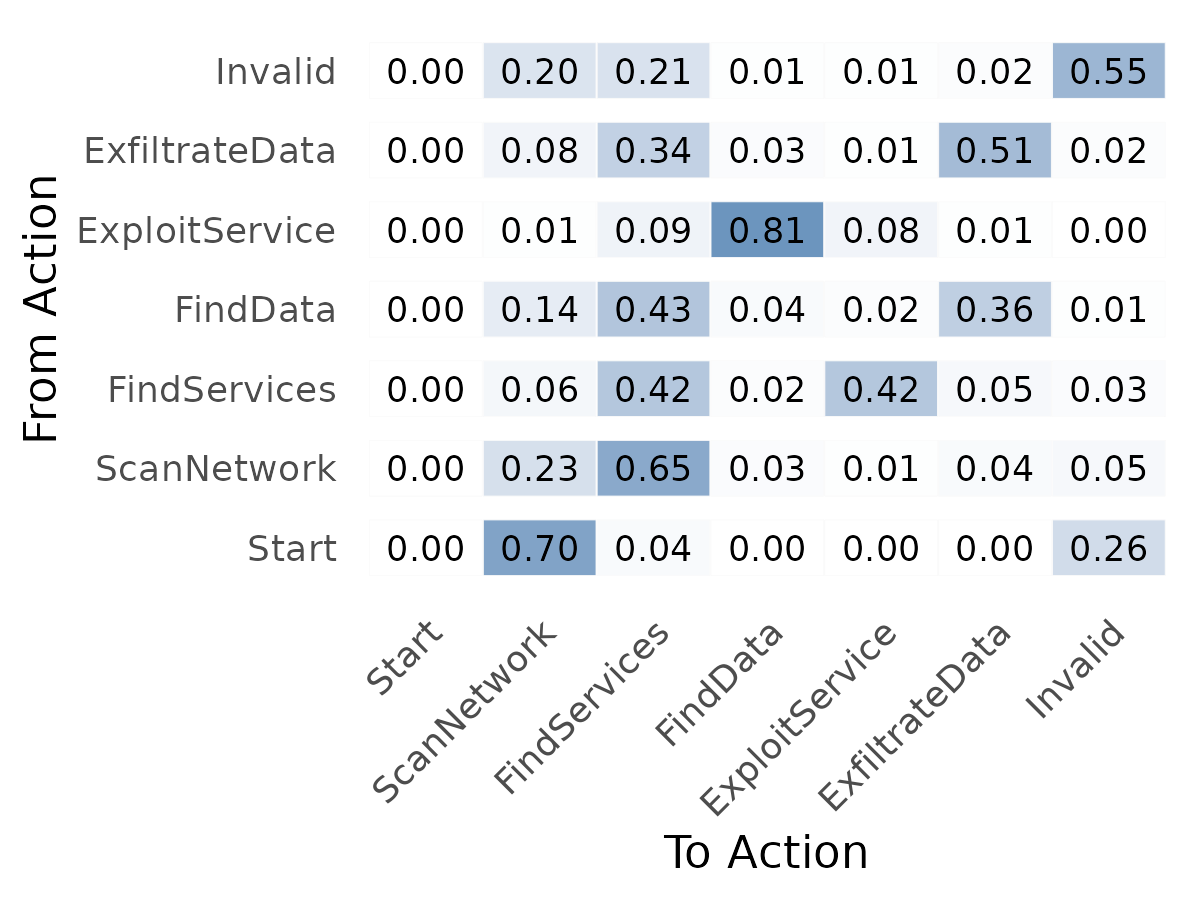}
        \caption{Transition Matrix}
        \label{fig:trans-matrix-zephyr-ft}
    \end{subfigure}
    \begin{subfigure}[b]{0.45\textwidth}
        \centering
        \includegraphics[trim={5cm 3cm 3cm 2.5cm},width=0.8\textwidth]{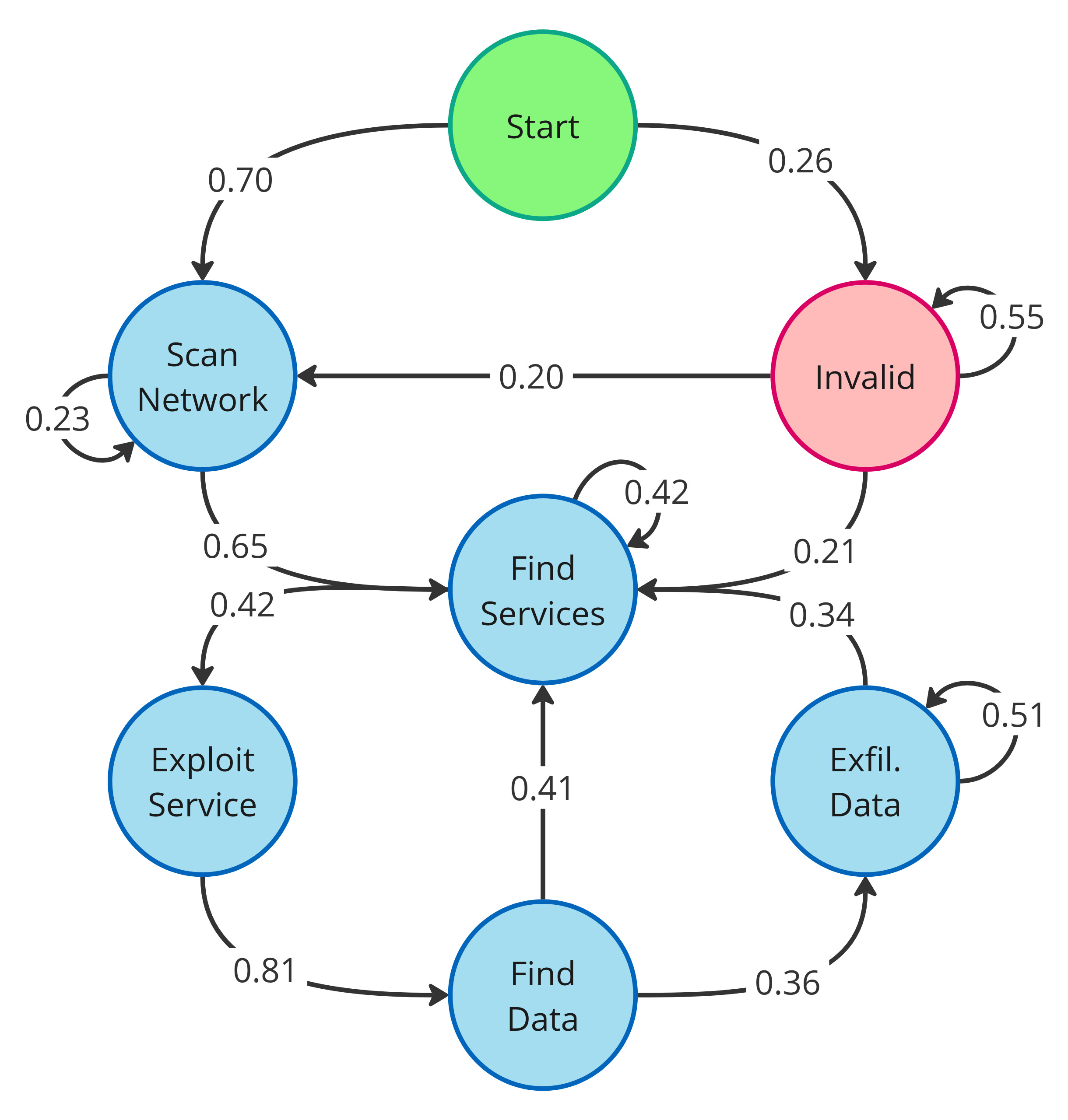}
        \caption{Key Action Transitions}
        \label{fig:key-action-zephyr-ft}
    \end{subfigure}
    \caption{Hackphyr action transitions}
    \label{fig:behavior-zephyr-ft}
\end{figure}

When comparing the behavior of the Hackphyr-based agent with the agent using GPT-4 (Figure~\ref{fig:behavior-zephyr-ft}), the transition matrix shows the agent is not as confident about taking the next actions. The next action transition distribution is more evenly distributed in all cases except the \texttt{ExploitService} and the \texttt{ScanNetwork}. However, when considering the key action transaction graph~\ref{fig:key-action-zephyr-ft}, we can observe the agent also follows a similar behavior pattern of systematic reconnaissance, service discovery, exploitation, data finding, and exfiltration. However, the Hackphyr agent seems to follow a slightly different strategy when scanning networks. In most episodes, the agent tried to scan all the networks before continuing to search for services. This situation is represented by the self-loop transition in the \texttt{ScanNetwork} node.

Another observable difference with the GPT-4-based agent is the transition to invalid actions from the \texttt{Start}. Once the agent has generated an invalid action, the most probable transition is to another invalid action  (with a probability of 0.55). The initial environment state information could be harder for the Hackphyr agent to understand. In addition, the lack of short-term memory information during the start could also cause the transition to invalid actions. 

\begin{figure}[!t]
    \centering
    \begin{subfigure}[b]{0.45\textwidth}
        \centering
        \includegraphics[width=1.05\textwidth]{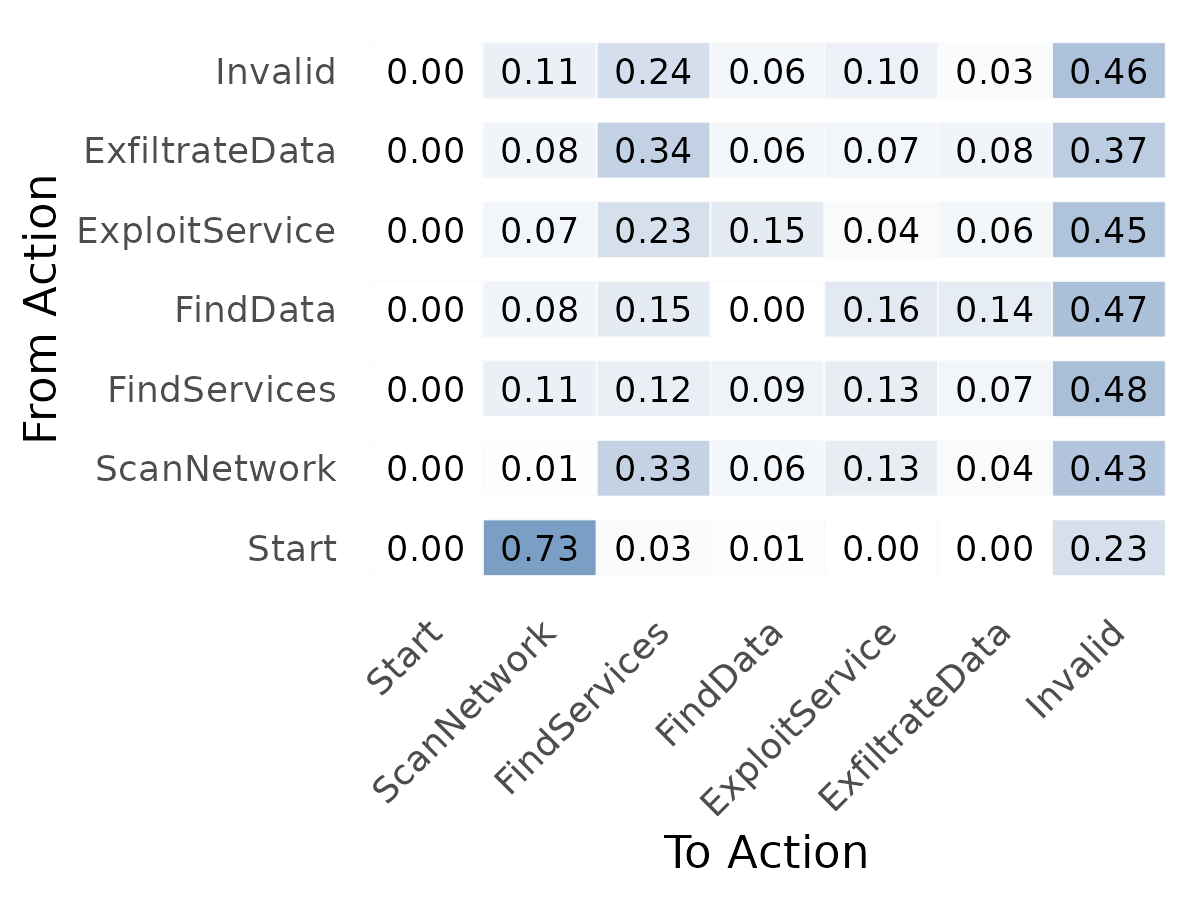}
        \caption{Transition Matrix}
        \label{fig:trans-matrix-zephyr}
    \end{subfigure}
    \begin{subfigure}[b]{0.45\textwidth}
        \centering
        \includegraphics[trim={5cm 3cm 3cm 2.5cm},width=0.85\textwidth]{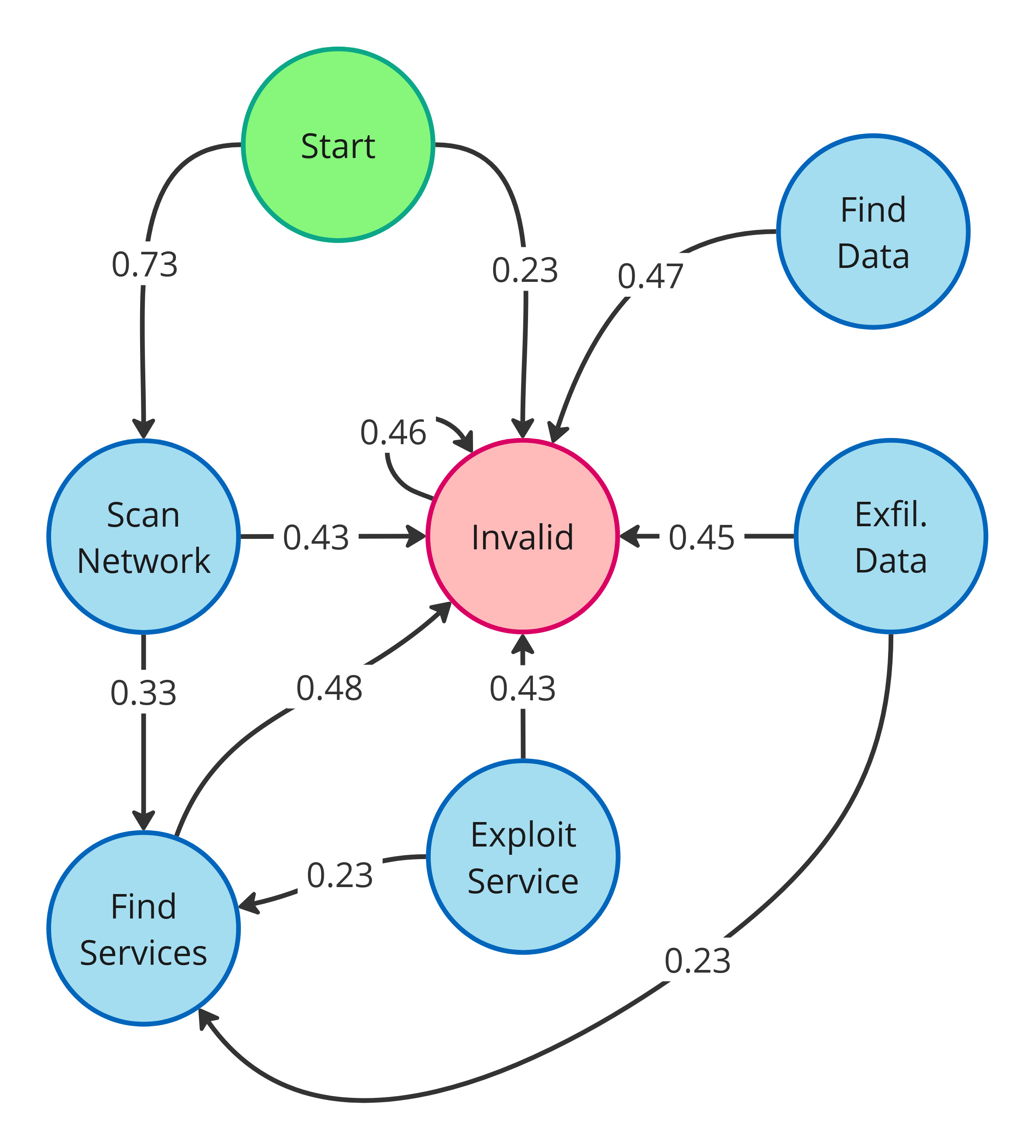}
        \caption{Key Action Transitions}
        \label{fig:key-action-zephyr}
    \end{subfigure}
    \caption{Zephyr-7b-$\beta$ action transitions}
    \label{fig:behavior-zephyr}
\end{figure}

Finally, Figure~\ref{fig:behavior-zephyr} shows the agent's behavior when using the base Zephyr-7b-$beta$ model. The transition matrix (Figure~\ref{fig:trans-matrix-zephyr}) shows a clear difference in the distribution of the action transition from GPT-4 and Hackphyr. The main difference is the high transition probability from any other possible action to an \textit{Invalid} action. In addition, it seems the agent is biased to transition to the \textit{FindServices} action, no matter the previous action.

The key actions transition graph in Figure~\ref{fig:key-action-zephyr} shows the lack of a clear behavior pattern for conducting the penetration testing techniques. At the \texttt{Start}, follows a logical \texttt{ScanNetwork} action, and the transition to \texttt{FindServices}. However, the required transitions from \texttt{FindeServices} to \texttt{ExploitService} and \textit{FindData} are not shown in the graph since they have a very low probability. Similar is the case of the transition from \texttt{FindData} to \texttt{ExfiltrateData}. Therefore, beyond scanning the network and finding services,  the agent cannot follow the proper action sequence for exfiltrating the data.  

In addition, the graph also shows the central role of the Invalid action node. All possible actions transition to an \textit{Invalid} one, with a considerable probability (between 0.25 and 0.48).

\section{Dataset Ablation Study}
\label{sec:ablation}

This ablation study aimed to determine which part of the dataset contributes more to the performance of the Hackphyr model during the SFT process. 


Since the dataset consists of three parts (see Section~\ref{sec:dataset-creation}), we constructed three different datasets where each one has one of the dataset parts removed:

\begin{itemize}
    \item Dataset UV contains parts I+II, with a focus on Understanding and Valid actions and 1526 samples
    \item Dataset UG, contains parts I+III, with a focus on Understanding and Good actions and 1189 samples
    \item Dataset VG contains parts II+III, with a focus on Valid and Good actions) and 563 samples
\end{itemize}

Each dataset was used to fine-tune a new model with the exact hyper-parameter settings described in Section~\ref{sec:best-model-selection}. We used the "full" scenario with and without the stochastic defender to compare their performance.


First, we analyze the effectiveness of the actions taken by the agents when using different dataset parts. The actions are categorized as \texttt{good}, \texttt{valid}, or \texttt{invalid}: a good action is performed by the agent that increases the current state; a valid action is appropriate and permissible in the current state, maintaining system integrity or preparing for future beneficial actions, but does not modify the state immediately; and an invalid action is not permissible in the current state and should be minimized by the agent to prevent errors and inefficiencies.

\begin{figure*}[t]
    \centering
    \includegraphics[width=1\linewidth]{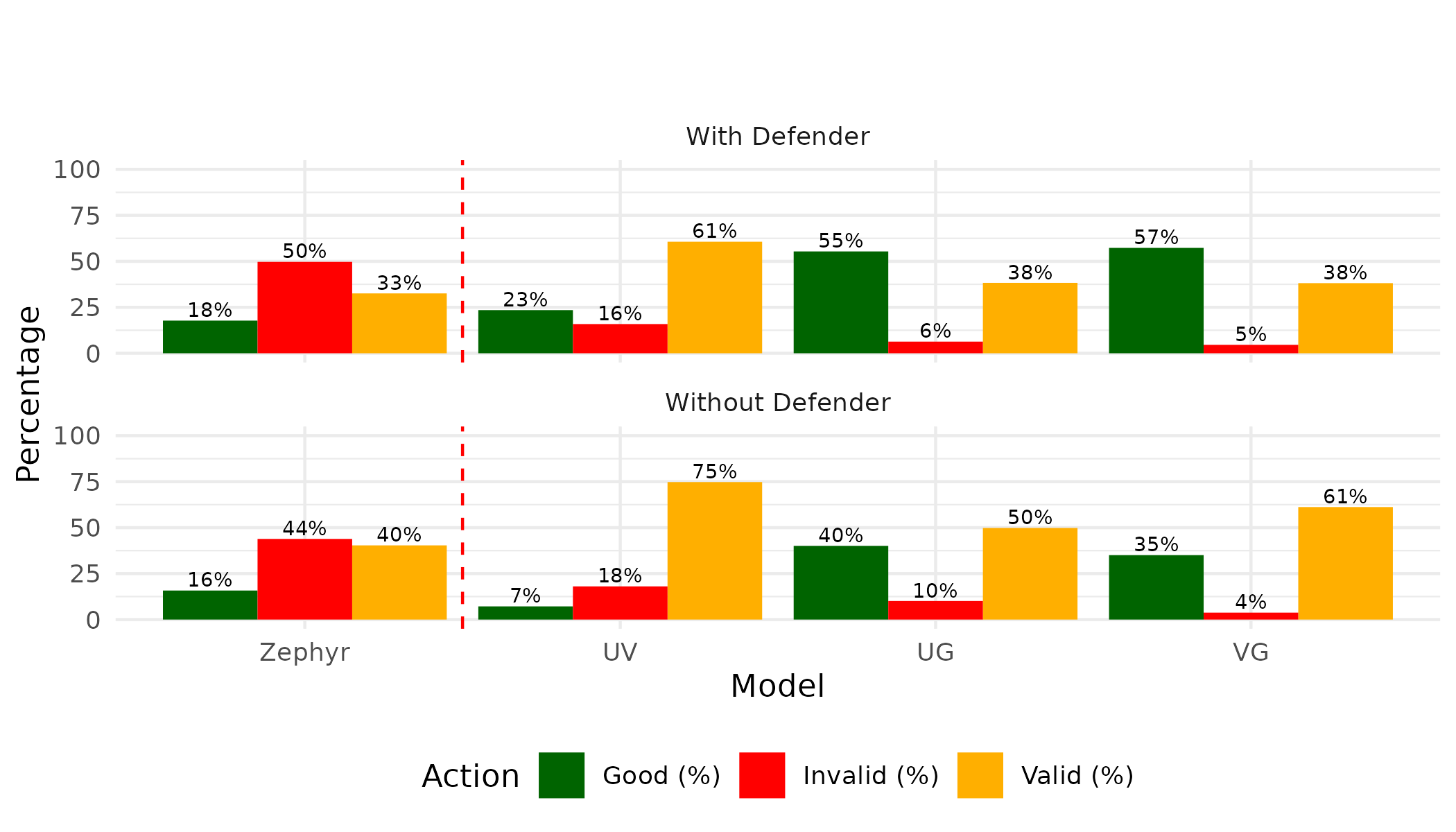}
    \caption{Percentage of good, valid, and invalid actions taken by agents trained with different dataset parts.}
    \label{fig:comparison-ablation-actions}
\end{figure*}

Figure~\ref{fig:comparison-ablation-actions} exhibits the percentage of each type of action each agent takes on the full scenario (with and without a defender) after 150 episodes. In addition, we provide the results for the agent using the base Zephyr-7b-$\beta$ without any fine-tuning process.

In the scenarios without the stochastic defender, Dataset UV does not show an observable increment in the number of good actions (7\%); however, the number of invalid actions is considerably reduced from 44\% to 18\%. Dataset UG shows a considerable increase in the number of good actions (40\%), and invalid actions are reduced to 10\%. Finally, in Dataset VG, good actions are slightly below those in Dataset UV (35\%), but the number of invalid actions is reduced to 4\%.

Similarly, in the case of the stochastic defender, all agents show improvements compared to the base Zephyr agent. Training with Dataset UV has an increase in the good action rate of 23\% (up from Zephyr's 18\%), with invalid actions reduced to 16\% (down from 50\%) and valid actions increased to 61\% (up from 33\%). Both agents trained with Dataset UG and Dataset VG  demonstrate significant improvements, behaving almost the same with good action rates of 55\% and 57\%, respectively, a drastic reduction in invalid actions to 6\% and 5\%, and an increase in valid actions to 38\%. 

In both scenarios (with and without the stochastic defender), the benefits provided by the different parts of the dataset are clear. In particular, the best results are observed on agents trained with UG and VG datasets, with a considerable increase in the number of good actions. In the case of dataset UV, despite reducing the number of invalid actions, the number of good actions remains low. 

\begin{figure*}[t]
    \centering
    \begin{subfigure}[b]{0.45\linewidth}
        \centering
        \includegraphics[width=1\linewidth,trim={0cm 1cm 0cm 0cm},clip]{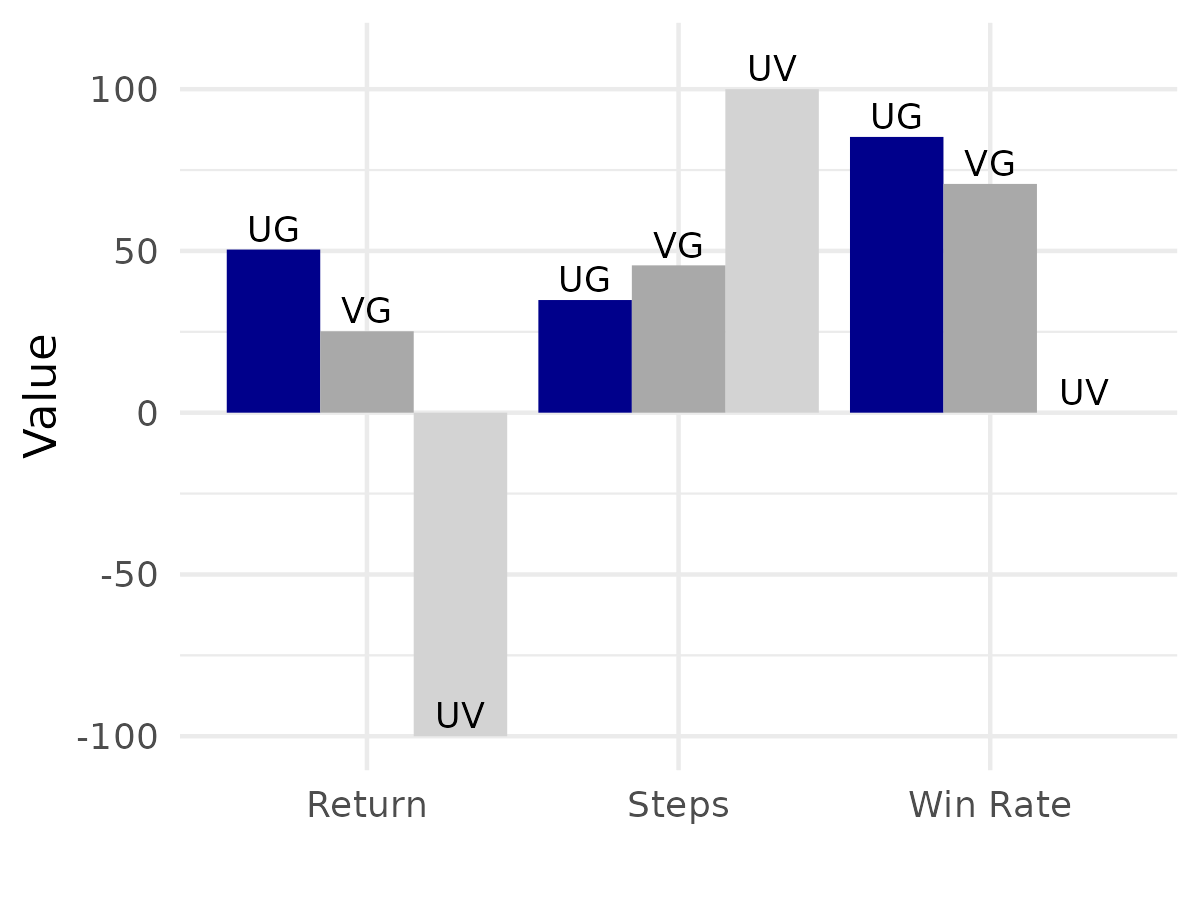}
        \caption{Without Stochastic Defender}
        \label{fig:without-defender}
    \end{subfigure}
    \hfill
    \begin{subfigure}[b]{0.45\linewidth}
        \centering
        \includegraphics[width=1\linewidth,trim={0cm, 1cm 0cm 0cm},clip]{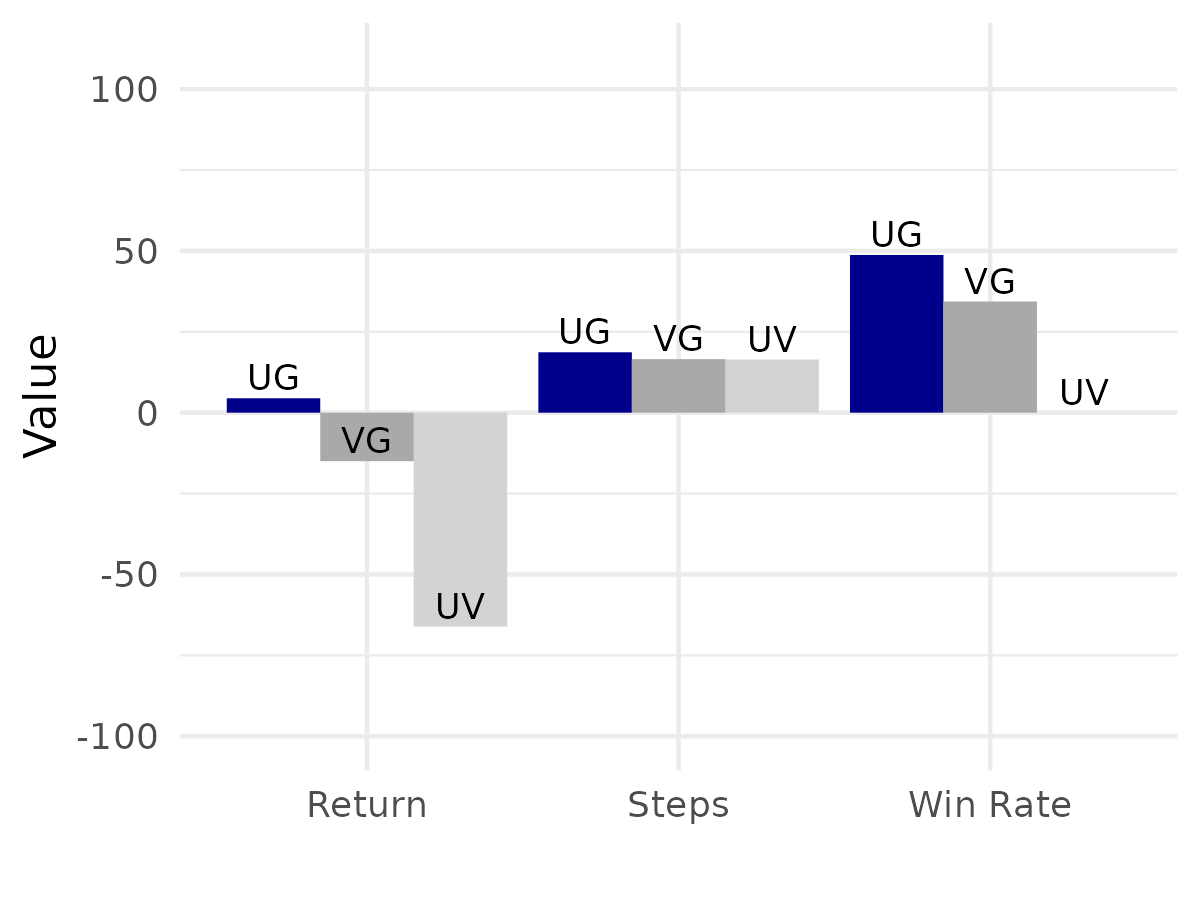}
        \caption{With Stochastic Defender}
        \label{fig:with-defender}
    \end{subfigure}
    
    \caption{Comparison of ablated agents in the large scenario with and without the stochastic defender.}
    \label{fig:comparison-ablation}
\end{figure*}

Being capable of taking valid actions is a necessary condition for an agent to be able to interact with the environment. However, that could not be enough to solve it. To illustrate the agents' performance, Figure~\ref{fig:comparison-ablation} presents the agents' results in terms of average return and average steps per episode when using the models across different datasets, along with the win rate over the 150 episodes. The model with the best performance in each metric is highlighted in blue. Note that higher values are better for return and win rate, whereas lower values are preferable for steps.

When the stochastic defender is not present (see Figure~\ref{fig:comparison-ablation}(a)), the performance of the agent fine-tuned with dataset UV is worse in every metric. The agent has not been able to win in any of the 150 episodes (0\% Win Rate). Consequently, the average return was -100, which means that the agent has exhausted the maximum number of steps allowed in the environment. So, despite the increase in the number of valid actions, the agent cannot generate enough good actions to explore and solve the environment. On the other hand, the fine-tuned version of agents using datasets UG and VG can solve the environment. In particular, the agent fine-tuned in dataset UG has shown an 85\% win rate over the 150 episodes with an average return of around 50.

Similarly, in the scenario with the stochastic defender (see subfigure~\ref{fig:with-defender}), the agent fine-tuned with dataset UV has not been able to win in any of the 150 episodes (0\% win rate). The fine-tuned version of agents with the other two datasets can solve the environment. In particular, agents using dataset UG have shown a 50\% win rate over the 150 episodes with an average return of around 5.


When the agent uses only Zephyr pre-trained data, there are many invalid actions (see Section~\ref{sec:results} for the results of the base Zephyr-7b-$\beta$ agent). This can be a cause for a low win rate in all scenarios. 

To improve the agents' performance, Part III is essential. Datasets using part III have shown a higher number of good actions. In addition, the combination of Part I and Part III shows the best results regarding win rate and average return. Meanwhile, the combination of Part II and Part III also performs well, though slightly less effectively. The differences in performance between these datasets might be attributed to the varying sizes of the datasets.

Finally, a high number of good actions does not necessarily imply the best agent. For instance, in the case of the scenario with the defender, the agent with Dataset VG shows the highest number of good actions, but in terms of win rate, the results are not as good as the agent using Dataset UG. Good actions are useful for exploring the environment, but sometimes it is not enough to win within the allowed maximum steps. 

\section{Discussion}
\label{sec:discussion}

Our fine-tuned model performed comparatively close to the capabilities of GPT-4, which has at least two orders of magnitude more parameters. While larger models like an 80 billion parameter model are available and may have potentially superior performance, they would require significant computational resources to train and deploy. This could limit their practical application on commercial hardware for specialized tasks.

Moreover, even a 7 billion parameter model, such as Hackphyr, requires a strong GPU to run inference. Quantization techniques hold promise as a potential solution by reducing the memory requirements of the model. However, quantization often comes at the cost of performance degradation. Further investigation is needed into alternative methods that balance computational efficiency with maintainable performance.

In this work, we used the same prompt across all models and agents to compare our agents fairly. However, different language models may respond better to tailored prompts. Future work will focus on optimizing our prompts using techniques such as DSPy~\cite{khattab2023dspy} to adapt them to the specific strengths of each model.

Finally, we focused primarily on attacking agents within the NetSecGame environment, mainly as a means of comparison to prior work. However, the principles are highly transferable to designing and implementing attacking and defending agents within varied reinforcement learning environments. Using specialized agents collaborating in multi-agent deployments, either with or without human involvement, is an exciting area for future work. We anticipate this line of work will contribute valuable insights regarding defense strategies against malicious agents and hope to extend it further in the future.

\section{Conclusions}
\label{sec:conclusions}

While powerful commercial language models can handle many tasks, there are often reasons that prohibit their use. In the security domain, companies are hesitant to share the data of their internal networks, for example. A variety of smaller models have been released under permissive licenses. These models require much less hardware to run, and they can be used on-premises. However, they are not as capable as the much larger commercial models. In this work, we fine-tuned a 7 billion parameter model to function as an attacking agent within a network security environment. The resulting model, Hackphyr, had a performance comparable to that of the most powerful commercial model (GPT-4) and outperformed other language models (GPT-3.5-turbo, Zephyr-7b-$beta$) and baselines (Q-learning agent) even in the most complex previously unseen scenario.

To fine-tune the base model, we generated a new dataset that aimed to address the weaknesses of the base model with regard to environment understanding and the generation of valid and useful actions at a given state. Finally, we performed an extensive analysis of each agent's actions that gave us insight into their behavior and shortcomings. This type of analysis can be used as a first step to explain the LLM agents' behavior and can be a useful tool in the future for all types of agents in similar environments.

\section{Acknowledgements}
The authors acknowledge support by the Strategic Support for the Development of Security Research in the Czech Republic 2019--2025 (IMPAKT 1) program, by the Ministry of the Interior of the Czech Republic under No. VJ02010020 -- AI-Dojo: Multi-agent testbed for the research and testing of AI-driven cyber security technologies.

\appendix
\section{Dataset examples}
\label{sec:appendix_dataset}

\subsection{Part I}
\label{subsec:appendix_dataset_partI}
Question:

\begin{tcolorbox}[colback=gray!5!white,colframe=gray!75!black]
\begin{small}
\begin{verbatim}
This is the current status of a computer network during a 
penetration testing procedure:
```
Known networks are 1.1.1.0/24
Known hosts are 2.2.2.3
Controlled hosts are 2.2.2.2 and 1.1.1.2
Known data for source host 1.1.1.2: are (User1, WebData)
Known services for host 1.1.1.1 are openssh
```
The actions you can take are ScanServices, FindData, ScanNetwork,
ExploitService, and ExfiltrateData.
The rules are:
{{rules}}

Can ScanNetwork action be used on a subnet that is not within
the known networks?

\end{verbatim}
\end{small}
\end{tcolorbox}

Answer:

\begin{tcolorbox}[colback=gray!5!white,colframe=gray!75!black]
\begin{small}
\begin{verbatim}
No, according to the rules, you can only scan networks that are
in your list of known networks.
\end{verbatim}
\end{small}
\end{tcolorbox}

\subsection{Part II}
\label{subsec:appendix_dataset_partII}
Question:

\begin{tcolorbox}[colback=gray!5!white,colframe=gray!75!black]
\begin{small}
\begin{verbatim}
You are a competent pentester and you are trying to teach others.
This is the current status of a computer network during a
penetration testing procedure:
```
Known networks are 1.1.1.0/24 and 2.2.2.0/24 
Known hosts are 2.2.2.3 and 2.2.2.4 and 1.1.1.1
Controlled hosts are 2.2.2.2 and 1.1.1.2 
Known data for source host 1.1.1.2: are (User1, WebData) 
Known services for host 1.1.1.1 are openssh
Know services for host 1.1.1.2 are openssh and http
Know services for host 2.2.2.3 are http and ldap
Know services for host 2.2.2.4 are openssh and http and ldap
```
The actions you can take are ScanServices, FindData, ScanNetwork,
ExploitService, and ExfiltrateData.
The rules are:
{{rules}}

Here is an example of each action in JSON format:
...

Provide all the ExploitService actions that can be taken in
the current status in the correct format.
\end{verbatim}
\end{small}
\end{tcolorbox}

Answer:
\begin{tcolorbox}[colback=gray!5!white,colframe=gray!75!black]
\begin{small}
\begin{verbatim}
Based on the current status and the rules provided, we
can perform ScanNetwork actions on the known networks.
Here are the ScanNetwork actions in the correct JSON format for
each of the known networks:

```json
{"action":"ScanNetwork", "parameters":
    {"target_network": "10.10.0.0/24"}}
{"action":"ScanNetwork", "parameters":
    {"target_network": "10.10.1.0/24"}}
{"action":"ScanNetwork", "parameters":
    {"target_network": "10.10.2.0/24"}}
```
\end{verbatim}
\end{small}
\end{tcolorbox}

\subsection{Part III}
\label{subsec:appendix_dataset_partIII}
Question:
\begin{tcolorbox}[colback=gray!5!white,colframe=gray!75!black]
\begin{small}
\begin{verbatim}
You are a pentester and you want to exfiltrate data
to target host 42.86.112.78.
The actions you can take are ScanServices, FindData, ScanNetwork,
ExploitService, and ExfiltrateData. 
The rules are:
{{rules}}

Current status:
Controlled hosts are 10.0.1.76 and 42.86.112.78
Known networks are 10.0.1.0/24 and 10.0.2.0/24 and 10.0.0.0/24
Known hosts are 
Known services are none
Known data are none

Here are some examples of actions:
{{examples}}

Based on the current status provided, here's a breakdown of
the objects and the actions that can be applied to them:

### Controlled Hosts:
- **10.0.1.76** and **42.86.112.78**
  - Actions:
    - **FindData**: You can find data in hosts you control.

### Known Networks:
- **10.0.1.0/24**, **10.0.2.0/24**, and **10.0.0.0/24**
  - Actions:
    - **ScanNetwork**: You can scan networks in your list of
    known networks to discover hosts.

### Known Hosts:
- None listed, but you can discover hosts by scanning the
known networks.

### Known Services:
- None listed, but you can discover services by scanning
the hosts you know.

### Known Data:
- None listed, but you can find data in the controlled hosts.

Provide the best next action in the correct JSON format. Action:
\end{verbatim}
\end{small}
\end{tcolorbox}

Answer:
\begin{tcolorbox}[colback=gray!5!white,colframe=gray!75!black]
\begin{small}
\begin{verbatim}
{"action": "ScanNetwork","parameters":
    {"target_network": "10.0.1.0/24"}}
\end{verbatim}
\end{small}
\end{tcolorbox}
\bibliographystyle{elsarticle-num}
\bibliography{ref2}





\end{document}